\documentclass[10pt,journal,compsoc]{IEEEtran}
\usepackage{graphicx}
\usepackage{balance}
\usepackage{subcaption}
\usepackage{cite}

\usepackage[usenames,dvipsnames]{color}
\usepackage{colortbl}
\definecolor{mygray}{gray}{.8}

\usepackage{algorithm}
\usepackage{algorithmic}

\usepackage{soul}
\usepackage{epstopdf}
\usepackage{hyperref}
\usepackage{multirow}
\usepackage{listings}
\usepackage{fancybox}
\usepackage{graphicx}
\usepackage{subcaption}
\usepackage{paralist}
\usepackage{ragged2e}
\usepackage{enumitem}
\usepackage{booktabs}
\usepackage{longtable}

\usepackage[framemethod=TikZ]{mdframed}
\usepackage{tcolorbox}
\makeatletter
\newcommand{\mybox}[1]{%
  \setbox0=\hbox{#1}%
  \setlength{\@tempdima}{\dimexpr\wd0+13pt}%
  \begin{tcolorbox}[boxrule=0.5pt, colback=white, arc=4pt,
      left=6pt,right=6pt,top=6pt,bottom=6pt,boxsep=0pt]
    #1
  \end{tcolorbox}
}
\usepackage{tikz}
\usepackage{pgfplots}
\usetikzlibrary{pgfplots.statistics,calc}
\usepackage{multicols}
\usepackage{color}
\usepackage{xcolor}
\usepackage{array}
\usepackage{amsmath}
\usepackage{centernot}
\usepackage{xspace}
\usepackage{url}
\usepackage{verbatim}
\usepackage{wrapfig}
\usepackage{tabularx}
\usepackage{ulem}
\usepackage{CJKulem}
\usepackage{amssymb}
\usepackage{mathrsfs}
\usepackage{MnSymbol,bbding,pifont}
\usepackage{makecell}

\newcommand{\papers}{102}

\usepackage[figuresright]{rotating}

\newcolumntype{s}{>{\hsize=.12\hsize}X}
\newcolumntype{f}{>{\hsize=.88\hsize}X}

\begin{document}

\title{Software Testing with Large Language Models: \\ Survey, Landscape,  and Vision}

\author{Junjie~Wang, Yuchao~Huang, Chunyang~Chen, Zhe~Liu, Song~Wang, Qing~Wang
\IEEEcompsocitemizethanks{\IEEEcompsocthanksitem J. Wang,Y. Huang, Z. Liu, Q. Wang are with State Key Laboratory of Intelligent Game, Institute of Software Chinese Academy of Sciences, and University of Chinese Academy of Sciences, Beijing, China. J. Wang and Q. Wang are corresponding authors.\protect\\
E-mail: \{junjie, yuchao2019, liuzhe2020, wq\}@iscas.ac.cn
\IEEEcompsocthanksitem C. Chen is with Monash University, Melbourne, Australia \protect\\
E-mail: chunyang.chen@monash.edu
\IEEEcompsocthanksitem S. Wang is with York University, Toronto, Canada. \protect\\
E-mail: wangsong@yorku.ca
}}


\IEEEtitleabstractindextext{%
\justify
\begin{abstract}
Pre-trained large language models (LLMs) have recently emerged as a breakthrough technology in natural language processing and artificial intelligence, with the ability to handle large-scale datasets and exhibit remarkable performance across a wide range of tasks. 
Meanwhile, software testing is a crucial undertaking that serves as a cornerstone for ensuring the quality and reliability of software products. 
As the scope and complexity of software systems continue to grow, the need for more effective software testing techniques becomes increasingly urgent, making it an area ripe for innovative approaches such as the use of LLMs.
This paper provides a comprehensive review of the utilization of LLMs in software testing. 
It analyzes {\papers} relevant studies that have used LLMs for software testing, from both the software testing and LLMs perspectives. 
The paper presents a detailed discussion of the software testing tasks for which LLMs are commonly used, among which test case preparation and program repair are the most representative. 
It also analyzes the commonly used LLMs, the types of prompt engineering that are employed, as well as the accompanied techniques with these LLMs.
It also summarizes the key challenges and potential opportunities in this direction. 
This work can serve as a roadmap for future research in this area, highlighting potential avenues for exploration, and identifying gaps in our current understanding of the use of LLMs in software testing. 
\end{abstract}

}

\maketitle


\IEEEpeerreviewmaketitle
\section{Introduction}
\label{sec_introduction}

Software testing is a crucial undertaking that serves as a cornerstone for ensuring the quality and reliability of software products. 
Without the rigorous process of software testing, software enterprises would be reluctant to release their products into the market, knowing the potential consequences of delivering flawed software to end-users.
By conducting thorough and meticulous testing procedures, software enterprises can minimize the occurrence of critical software failures, usability issues, or security breaches that could potentially lead to financial losses or jeopardize user trust. Additionally, software testing helps to reduce maintenance costs by identifying and resolving issues early in the development lifecycle, preventing more significant complications down the line \cite{Myers2004artofsoftwaretesting,pezze2007softwaretesting}.

The significance of software testing has garnered substantial attention within the research and industrial communities. In the field of software engineering, it stands as an immensely popular and vibrant research area. One can observe the undeniable prominence of software testing by simply examining the landscape of conferences and symposiums focused on software engineering. Amongst these events, topics related to software testing consistently dominate the submission numbers and are frequently selected for publication.

While the field of software testing has gained significant popularity, there remain dozens of challenges that have not been effectively addressed. For example, one such challenge is automated unit test case generation. 
Although various approaches, including search-based~\cite{harman2010theoretical,pedro2023interactive}, constraint-based~\cite{xiao2013characteristics} or random-based~\cite{pacheco2007feedback} techniques to generate a suite of unit tests, the coverage and the meaningfulness of the generated tests are still far from satisfactory \cite{64noUnitTest,123ChatgptVsSbst}.
Similarly, when it comes to mobile GUI testing,  existing studies with random-/rule-based methods \cite{Monkey,li2017droidbot}, model-based methods \cite{su2017guided,dong2020time}, and learning-based methods \cite{pan2020reinforcement} are unable to understand the semantic information of the GUI page and often fall short in achieving comprehensive coverage \cite{60GUITesting,su2021benchmarkingGuitesting}.
Considering these limitations, numerous research efforts are currently underway to explore innovative techniques that can enhance the efficacy of software testing tasks, among which large language models are the most promising ones.

Large language models (LLMs) such as T5 and GPT-3 have revolutionized the field of natural language processing (NLP) and artificial intelligence (AI). 
These models, initially pre-trained on extensive corpora, have exhibited remarkable performance across a wide range of NLP tasks including question-answering, machine translation, and text generation \cite{Shanahan2022Talking,zhao2023surveyLLM,Kojima2022Large,wei2022chainOfThought}. 
In recent years, there has been a significant advancement in LLMs with the emergence of models capable of handling even larger-scale datasets. 
This expansion in model size has not only led to improved performance but also opened up new possibilities for applying LLMs as Artificial General Intelligence.
Among these advanced LLMs, models like ChatGPT\footnote{https://openai.com/blog/chatgpt} and LLaMA\footnote{https://ai.meta.com/blog/large-language-model-llama-meta-ai/} boast billions of parameters. 
Such models hold tremendous potential for tackling complex practical tasks in domains like code generation and artistic creation.
With their expanded capacity and enhanced capabilities, LLMs have become game-changers in NLP and AI, and are driving advancements in other fields like coding and software testing.

LLMs have been used for various coding-related tasks including code generation and code recommendation~\cite{CodeGeneration1, CodeGeneration2, CodeGeneration3,dong2023selfcollaboration}.
On one hand, in software testing, there are many tasks related to code generation, such as unit test generation \cite{64noUnitTest}, where the utilization of LLMs is expected to yield good performance. 
On the other hand, software testing possesses unique characteristics that differentiate it from code generation. For example, code generation primarily focuses on producing a single, correct code snippet, whereas software testing often requires generating diverse test inputs to ensure better coverage of the software under test \cite{Myers2004artofsoftwaretesting}. 
The existence of these differences introduces new challenges and opportunities when employing LLMs for software testing. 
Moreover, people have benefited from the excellent performance of LLMs in generation and inference tasks, leading to the emergence of dozens of new practices that use LLMs for software testing.

\begin{figure}[t!]
\centering
\includegraphics[width=\linewidth]{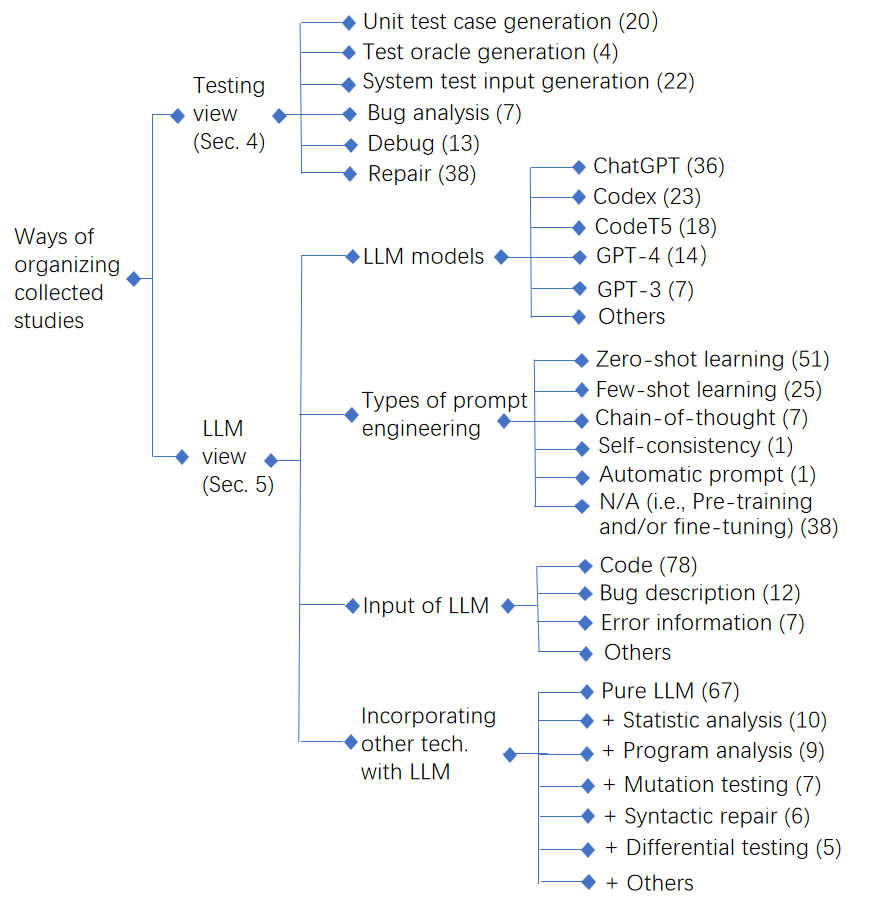}
\caption{Structure of the contents in this paper (the numbers in bracket indicates the number of involved papers, and a paper might involve zero or multiple items)}
\label{fig:overview}
\vspace{-0.1in}
\end{figure}

This article presents a comprehensive review of the utilization of LLMs in software testing. 
We collect {\papers} relevant papers and conduct a thorough analysis from both software testing and LLMs perspectives, as roughly summarized in Figure \ref{fig:overview}.

From the viewpoint of software testing, our analysis involves an examination of the specific software testing tasks for which LLMs are employed. 
Results show that LLMs are commonly used for test case preparation (including unit test case generation, test oracle generation, and system test input generation), program debugging, and bug repair, while we do not find the practices for applying LLMs in the tasks of early testing life-cycle (such as test requirement, test plan, etc).
For each test task, we would provide detailed illustrations showcasing the utilization of LLMs in addressing the task, highlighting commonly-used practices, tracking technology evolution trends, and summarizing achieved performance, so as to facilitate readers in gaining a thorough overview of how LLMs are employed across various testing tasks.

From the viewpoint of LLMs, our analysis includes the commonly used LLMs in these studies, the types of prompt engineering, the input of the LLMs, as well as the accompanied techniques with these LLMs.
Results show that about one-third of the studies utilize the LLMs through pre-training or fine-tuning schema, while the others employ prompt engineering to communicate with LLMs to steer their behavior for desired outcomes. 
For prompt engineering, the zero-shot learning and few-shot learning strategies are most commonly used, while other advances like chain-of-thought promoting and self-consistency are rarely utilized.  
Results also show that traditional testing techniques like differential testing and mutation testing are usually accompanied by LLMs to help generate more diversified tests. 

Furthermore, we summarize the key challenges and potential opportunities in this direction. 
Although software testing with LLMs has undergone significant growth in the past two years, there are still challenges in achieving high coverage of the testing, test oracle problem, rigorous evaluations, and real-world application of LLMs in software testing. 
Since it is a new emerging field, there are many research opportunities, including exploring LLMs in an early stage of testing, exploring LLMs for more types of software and non-functional testing, exploring advanced prompt engineering, as well as incorporating LLMs with traditional techniques. 


This paper makes the following contributions:

\begin{itemize}

\item We thoroughly analyze {\papers} relevant studies that used LLMs for software testing, regarding publication trends, distribution of publication venues, etc.

\item We conduct a comprehensive analysis from the perspective of software testing to understand the distribution of software testing tasks with LLM and present a thorough discussion about how these tasks are solved with LLM. 

\item We conduct a comprehensive analysis from the perspective of LLMs, and uncover the commonly-used LLMs, the types of prompt engineering, input of the LLMs, as well as the accompanied techniques with these LLMs.

\item We highlight the challenges in existing studies and present potential opportunities for further studies. 

\end{itemize}

We believe that this work will be valuable to both researchers and practitioners in the field of software engineering, as it provides a comprehensive overview of the current state and future vision of using LLMs for software testing. 
For researchers, this work can serve as a roadmap for future research in this area, highlighting potential avenues for exploration and identifying gaps in our current understanding of the use of LLMs in software testing.
For practitioners, this work can provide insights into the potential benefits and limitations of using LLMs for software testing, as well as practical guidance on how to effectively integrate them into existing testing processes.
By providing a detailed landscape of the current state and future vision of using LLMs for software testing, this work can help accelerate the adoption of this technology in the software engineering community and ultimately contribute to improving the quality and reliability of software systems.

\section{Background}
\label{sec_background}

\subsection{Large Language Model (LLM)}
\label{subsec_background_LLM}

Recently, pre-trained language models (PLMs) have been proposed by pretraining Transformer-based models over large-scale corpora, showing strong capabilities in solving various natural language processing (NLP) tasks \cite{Shanahan2022Talking,zhao2023surveyLLM,Kojima2022Large,wei2022chainOfThought}. Studies have shown that model scaling can lead to improved model capacity, prompting researchers to investigate the scaling effect through further parameter size increases. Interestingly, when the parameter scale exceeds a certain threshold, these larger language models demonstrate not only significant performance improvements but also special abilities such as in-context learning, which are absent in smaller models such as BERT. 

To discriminate the language models in different parameter scales, the research community has coined the term large language models (LLM) for the PLMs of significant size. 
LLMs typically refer to language models that have hundreds of billions (or more) of parameters and are trained on massive text data such as GPT-3, PaLM, Codex, and LLaMA. LLMs are built using the Transformer architecture, which stacks multi-head attention layers in a very deep neural network. Existing LLMs adopt similar model architectures (Transformer) and pre-training objectives (language modeling) as small language models, but largely scale up the model size, pre-training data, and total compute power. This enables LLMs to better understand natural language and generate high-quality text based on given context or prompts.

Note that, in existing literature, there is no formal consensus on the minimum parameter scale for LLMs, since the model capacity is also related to data size and total compute. 
In a recent survey of LLMs \cite{zhao2023surveyLLM}, the authors focus on discussing the language models with a model size larger than 10B.
Under their criteria, the first LLM is T5 released by Google in 2019, followed by GPT-3 released by OpenAI in 2020, and there are more than thirty LLMs released between 2021 and 2023 indicating its popularity. 
In another survey of unifying LLMs and knowledge graphs \cite{pan2023unifying}, the authors categorize the LLMs into three types: encoder-only (e.g., BERT), encoder-decoder (e.g., T5), and decoder-only network architecture (e.g., GPT-3). 
In our review, we take into account the categorization criteria of the two surveys and only consider the encoder-decoder and decoder-only network architecture of pre-training language models, since they can both support generative tasks. 
We do not consider the encoder-only network architecture because they cannot handle generative tasks, were proposed relatively early (e.g., BERT in 2018), and there are almost no models using this architecture after 2021.
In other words, the LLMs discussed in this paper not only include models with parameters of over 10B (as mentioned in \cite{zhao2023surveyLLM}) but also include other models that use the encoder-decoder and decoder-only network architecture (as mentioned in \cite{pan2023unifying}), such as BART with 140M parameters and GPT-2 with parameter sizes ranging from 117M to 1.5B.
This is also to potentially include more studies to demonstrate the landscape of this topic. 

\subsection{Software Testing}
\label{subsec_testLifecycle}



Software testing is a crucial process in software development that involves evaluating the quality of a software product. The primary goal of software testing is to identify defects or errors in the software system that could potentially lead to incorrect or unexpected behavior.
The whole life cycle of software testing typically includes the following tasks (demonstrated in Figure \ref{fig:testingTasks}):
\begin{itemize}
\item Requirement Analysis: analyze the software requirements and identify the testing objectives, scope, and criteria.
\item Test Plan: develop a test plan that outlines the testing strategy, test objectives, and schedule.

\item Test Design and Review: develop and review the test cases and test suites that align with the test plan and the requirements of the software application.

\item Test Case Preparation: the actual test cases are prepared based on the designs created in the previous stage. 

\item Test Execution: execute the tests that were designed in the previous stage. The software system is executed with the test cases and the results are recorded.

\item Test Reporting: analyze the results of the tests and generate reports that summarize the testing process and identify any defects or issues that were discovered.

\item Bug Fixing and Regression Testing: defects or issues identified during testing are reported to the development team for fixing. Once the defects are fixed, regression testing is performed to ensure that the changes have not introduced new defects or issues.

\item Software Release: once the software system has passed all of the testing stages and the defects have been fixed, the software can be released to the customer or end user.

\end{itemize}

The testing process is iterative and may involve multiple cycles of the above stages, depending on the complexity of the software system and the testing requirements.

During the testing phase, various types of tests may be performed, including unit tests, integration tests, system tests, and acceptance tests.

\begin{itemize}

\item Unit Testing involves testing individual units or components of the software application to ensure that they function correctly. 

\item Integration Testing involves testing different modules or components of the software application together to ensure that they work correctly as a system. 

\item System Testing involves testing the entire software system as a whole, including all the integrated components and external dependencies. 

\item Acceptance Testing involves testing the software application to ensure that it meets the business requirements and is ready for deployment. 

\end{itemize}

In addition, there can be functional testing, performance testing, unit testing, security testing, accessibility testing, etc, which explores various aspects of the software under test~\cite{myers2004art}. 
\section{Paper Selection and Review Schema}
\label{sec_review_method}




\begin{table*}[!t]
\caption{Details of the collected papers}
\label{tab:paperOverview}
\centering
\scriptsize
\scalebox{0.78}{
\begin{tabular}{@{}l|l|l|l|l@{}}
\toprule
\textbf{ID} & \textbf{Topic} & \textbf{Paper title} & \textbf{Year} & \textbf{Reference} \\ \midrule
1 & Unit test case generation & Unit Test Case Generation with Transformers and Focal Context & 2020 & \cite{28unitTest} \\
2 & Unit test case generation & Codet: Code Generation with Generated Tests & 2022 & \cite{18codetGeneratedTests} \\
3 & Unit test case generation & Interactive Code Generation via Test-Driven User-Intent Formalization & 2022 & \cite{17interactiveCodeGeneration} \\
4 & Unit test case generation & A3Test: Assertion-Augmented Automated Test Case Generation & 2023 & \cite{23a3testTestGeneration} \\
5 & Unit test case generation & An Empirical Evaluation of Using Large Language Models for Automated Unit   Test Generation & 2023 & \cite{25adaptiveTestGeneration} \\
6 & Unit test case generation & An Initial Investigation of ChatGPT Unit Test Generation Capability & 2023 & \cite{150AnInitialInvestigation} \\
7 & Unit test case generation & Automated Test Case Generation Using Code Models and Domain Adaptation & 2023 & \cite{119AutomatedTestCase} \\
8 & Unit test case generation & Automatic Generation of Test Cases based on Bug Reports: a Feasibility   Study with Large Language Models & 2023 & \cite{110AutomaticGenerationOf} \\
9 & Unit test case generation & Can Large Language Models Write Good Property-Based Tests? & 2023 & \cite{121CanLargeLanguage} \\
10 & Unit test case generation & CAT-LM Training Language Models on Aligned Code And Tests & 2023 & \cite{111CatLmTraining} \\
11 & Unit test case generation & ChatGPT vs SBST: A Comparative Assessment of Unit Test Suite Generation & 2023 & \cite{123ChatgptVsSbst} \\
12 & Unit test case generation & ChatUniTest: a ChatGPT-based Automated Unit Test Generation Tool & 2023 & \cite{63chatUniTest} \\
13 & Unit test case generation & CODAMOSA: Escaping Coverage Plateaus in Test Generation with Pre-trained   Large Language Models & 2023 & \cite{55codamosaTestGeneration} \\
14 & Unit test case generation & Effective Test Generation Using Pre-trained Large Language Models and   Mutation Testing & 2023 & \cite{117EffectiveTestGeneration} \\
15 & Unit test case generation & Exploring the Effectiveness of Large Language Models in Generating Unit   Tests & 2023 & \cite{66generatingUnitTests} \\
16 & Unit test case generation & How Well does LLM Generate Security Tests? & 2023 & \cite{112HowWellDoes} \\
17 & Unit test case generation & No More Manual Tests? Evaluating and Improving ChatGPT for Unit Test   Generation & 2023 & \cite{64noUnitTest} \\
18 & Unit test case generation & Prompting Code Interpreter to Write Better Unit Tests on Quixbugs   Functions & 2023 & \cite{113PromptingCodeInterpreter} \\
19 & Unit test case generation & Reinforcement Learning from Automatic Feedback for High-Quality Unit Test   Generation & 2023 & \cite{steenhoek2023reinforcement} \\
20 & Unit test case generation & Unit Test Generation using Generative AI: A Comparative Performance   Analysis of Autogeneration Tools & 2023 & \cite{bhatia2023unit} \\
21 & Test oracle generation & Generating Accurate Assert Statements for Unit Test Cases Using   Pretrained Transformers & 2022 & \cite{39assertStatements} \\
22 & Test oracle generation & Learning Deep Semantics for Test Completion & 2023 & \cite{24semanticsTestCompletion} \\
23 & Test oracle generation; Program repair & Using Transfer Learning for Code-Related Tasks & 2023 & \cite{80UsingTransferLearning} \\
24 & Test oracle generation; Program repair & Retrieval-Based Prompt Selection for Code-Related Few-Shot Learning & 2023 & \cite{56retrievalPromptSelection} \\
25 & System test input generation & Automated Conformance Testing for JavaScript Engines via Deep Compiler   Fuzzing & 2021 & \cite{14conformanceTesting} \\
26 & System test input generation & Fill in the Blank: Context-aware Automated Text Input Generation for   Mobile GUI Testing & 2022 & \cite{26fillBlank} \\
27 & System test input generation & Large Language Models are Pretty Good Zero-Shot Video Game Bug Detectors & 2022 & \cite{7gameBugDetectors} \\
28 & System test input generation & Slgpt: Using Transfer Learning to Directly Generate Simulink Model Files   and Find Bugs in the Simulink Toolchain & 2021 & \cite{12slgptSimulink} \\
29 & System test input generation & Augmenting Greybox Fuzzing with Generative AI & 2023 & \cite{141AugmentingGreyboxFuzzing} \\
30 & System test input generation & Automated Test Case Generation Using T5 and GPT-3 & 2023 & \cite{143AutomatedTestCase} \\
31 & System test input generation & Automating GUI-based Software Testing with GPT-3 & 2023 & \cite{84automatingGUI} \\
32 & System test input generation & AXNav: Replaying Accessibility Tests from Natural Language & 2023 & \cite{159AxnavReplayingAccessibility} \\
33 & System test input generation & Can ChatGPT Advance Software Testing Intelligence? An Experience Report   on Metamorphic Testing & 2023 & \cite{106CanChatgptAdvance} \\
34 & System test input generation & Efficient Mutation Testing via Pre-Trained Language Models & 2023 & \cite{30mutationTesting} \\
35 & System test input generation & Large Language Models are Edge-Case Generators:Crafting Unusual Programs   for Fuzzing Deep Learning Libraries & 2023 & \cite{19fuzzDeepLearningLibraries} \\
36 & System test input generation & Large Language Models are Zero Shot Fuzzers: Fuzzing Deep Learning   Libraries via Large Language Models & 2023 & \cite{13fuzzDeepLearningLibraries} \\
37 & System test input generation & Large Language Models for Fuzzing Parsers (Registered Report) & 2023 & \cite{152LargeLanguageModels} \\
38 & System test input generation & LLM for Test Script Generation and Migration: Challenges, Capabilities,   and Opportunities & 2023 & \cite{115LlmForTest} \\
39 & System test input generation & Make LLM a Testing Expert: Bringing Human-like Interaction to Mobile GUI   Testing via Functionality-aware Decisions & 2023 & \cite{60GUITesting} \\
40 & System test input generation & PentestGPT: An LLM-empowered Automatic Penetration Testing Tool & 2023 & \cite{120PentestgptAnLlm} \\
41 & System test input generation & SMT Solver Validation Empowered by Large Pre-Trained Language Models & 2023 & \cite{204SmtSolverValidation} \\
42 & System test input generation & TARGET: Automated Scenario Generation from Traffic Rules for Testing   Autonomous Vehicles & 2023 & \cite{62targetTestGeneration} \\
43 & System test input generation & Testing the Limits: Unusual Text Inputs Generation for Mobile App Crash   Detection with Large Language Model & 2023 & \cite{107TestingTheLimits} \\
44 & System test input generation & Understanding Large Language Model Based Fuzz Driver Generation & 2023 & \cite{140UnderstandingLargeLanguage} \\
45 & System test input generation & Universal Fuzzing via Large Language Models & 2023 & \cite{166Fuzz4allUniversalFuzzing} \\
46 & System test input generation & Variable Discovery with Large Language Models for Metamorphic Testing of   Scientific Software & 2023 & \cite{82variablediscovery} \\
47 & System test input generation & White-box Compiler Fuzzing Empowered by Large Language Models & 2023 & \cite{175WhiteBoxCompiler} \\
48 & Bug analysis & Itiger: an Automatic Issue Title Generation Tool & 2022 & \cite{1itigerIssueTitle} \\
49 & Bug analysis & CrashTranslator: Automatically Reproducing Mobile Application Crashes   Directly from Stack Trace & 2023 & \cite{173CrashtranslatorAutomaticallyReproducing} \\
50 & Bug analysis & Cupid: Leveraging ChatGPT for More Accurate Duplicate Bug Report   Detection & 2023 & \cite{132CupidLeveragingChatgpt} \\
51 & Bug analysis & Employing Deep Learning and Structured Information Retrieval to Answer   Clarification Questions on Bug Reports & 2023 & \cite{134EmployingDeepLearning} \\
52 & Bug analysis & Explaining Software Bugs Leveraging Code Structures in Neural Machine   Translation & 2022 & \cite{5explainingSoftwareBugs} \\
53 & Bug analysis & Prompting Is All Your Need: Automated Android Bug Replay with Large   Language Models & 2023 & \cite{161PromptingIsAll} \\
54 & Bug analysis & Still Confusing for Bug-Component Triaging? Deep Feature Learning and   Ensemble Setting to Rescue & 2023 & \cite{145StillConfusingFor} \\
55 & Debug & Detect-Localize-Repair: A Unified Framework for Learning to Debug with   CodeT5 & 2022 & \cite{31detectLocalizeRepair} \\
56 & Debug & Large Language Models are Few-shot Testers: Exploring LLM-based General   Bug Reproduction & 2022 & \cite{9generalBugReproduction} \\
57 & Debug & A Preliminary Evaluation of LLM-Based Fault Localization & 2023 & \cite{163APreliminaryEvaluation} \\
58 & Debug & Addressing Compiler Errors: Stack Overflow or Large Language Models? & 2023 & \cite{171AddressingCompilerErrors} \\
59 & Debug & Can LLMs Demystify Bug Reports? & 2023 & \cite{130CanLlmsDemystify} \\
60 & Debug & Dcc --help: Generating Context-Aware Compiler Error Explanations with   Large Language Models & 2023 & \cite{164DccHelpGenerating} \\
61 & Debug & Explainable Automated Debugging via Large Language Model-driven   Scientific Debugging & 2023 & \cite{54explainableDebugging} \\
62 & Debug & Large Language Models for Test-Free Fault Localization & 2023 & \cite{168LargeLanguageModels} \\
63 & Debug & Large Language Models in Fault Localisation & 2023 & \cite{136LargeLanguageModels} \\
64 & Debug & LLM4CBI: Taming LLMs to Generate Effective Test Programs for Compiler Bug   Isolation & 2023 & \cite{122Llm4cbiTamingLlms} \\
65 & Debug & Nuances are the Key: Unlocking ChatGPT to Find Failure-Inducing Tests   with Differential Prompting & 2023 & \cite{67failureTestCases} \\
66 & Debug & Teaching Large Language Models to Self-Debug & 2023 & \cite{139TeachingLargeLanguage} \\
67 & Debug; Program repair & A study on Prompt Design, Advantages and Limitations of ChatGPT for Deep   Learning Program Repair & 2023 & \cite{52promptDesign} \\
68 & Program repair & Examining Zero-Shot Vulnerability Repair with Large Language Models & 2022 & \cite{35vulnerabilityRepair} \\
69 & Program repair & Automated Repair of Programs from Large Language Models & 2022 & \cite{34automatedRepair} \\
70 & Program repair & Fix Bugs with Transformer through a Neural-Symbolic Edit Grammar & 2022 & \cite{10neuralSymbolic} \\
71 & Program repair & Practical Program Repair in the Era of Large Pre-trained Language Models & 2022 & \cite{33practicalProgramRepair} \\
72 & Program repair & Repairing Bugs in Python Assignments Using Large Language Models & 2022 & \cite{8bugsPythonAssignments} \\
73 & Program repair & Towards JavaScript Program Repair with Generative Pre-trained Transformer   (GPT-2) & 2022 & \cite{38javaScriptProgram} \\
74 & Program repair & An Analysis of the Automatic Bug Fixing Performance of ChatGPT & 2023 & \cite{4analysisBugFixing} \\
75 & Program repair & An Empirical Study on Fine-Tuning Large Language Models of Code for   Automated Program Repair & 2023 & \cite{206AnEmpiricalStudy} \\
76 & Program repair & An Evaluation of the Effectiveness of OpenAI's ChatGPT for Automated   Python Program Bug Fixing using QuixBugs & 2023 & \cite{144AnEvaluationOf} \\
77 & Program repair & An Extensive Study on Model Architecture and Program Representation in   the Domain of Learning-based Automated Program Repair & 2023 & \cite{147AnExtensiveStudy} \\
78 & Program repair & Can OpenAI's Codex Fix Bugs? An Evaluation on QuixBugs & 2022 & \cite{44codexQuixBugs} \\
79 & Program repair & CIRCLE: Continual Repair Across Programming Languages & 2022 & \cite{40circleContinualRepair} \\
80 & Program repair & Coffee: Boost Your Code LLMs by Fixing Bugs with Feedback & 2023 & \cite{moon2023coffee} \\
81 & Program repair & Copiloting the Copilots: Fusing Large Language Models with Completion   Engines for Automated Program Repair & 2023 & \cite{201CopilotingTheCopilots} \\
82 & Program repair & Domain Knowledge Matters: Improving Prompts with Fix Templates for   Repairing Python Type Errors & 2023 & \cite{DomainKnowledge} \\
83 & Program repair & Enhancing Genetic Improvement Mutations Using Large Language Models & 2023 & \cite{167EnhancingGeneticImprovement} \\
84 & Program repair & FixEval: Execution-based Evaluation of Program Fixes for Programming   Problems & 2023 & \cite{149FixevalExecutionBased} \\
85 & Program repair & Fixing Hardware Security Bugs with Large Language Models & 2023 & \cite{36securityBugs} \\
86 & Program repair & Fixing Rust Compilation Errors using LLMs & 2023 & \cite{104FixingRustCompilation} \\
87 & Program repair & Framing Program Repair as Code Completion & 2022 & \cite{41repairAsCompletion} \\
88 & Program repair & Frustrated with Code Quality Issues? LLMs can Help! & 2023 & \cite{135FrustratedWithCode} \\
89 & Program repair & GPT-3-Powered Type Error Debugging: Investigating the Use of Large   Language Models for Code Repair & 2023 & \cite{151Gpt3Powered} \\
90 & Program repair & How Effective Are Neural Networks for Fixing Security Vulnerabilities & 2023 & \cite{57howEffective} \\
91 & Program repair & Impact of Code Language Models on Automated Program Repair & 2023 & \cite{32impactLanguageModels} \\
92 & Program repair & Inferfix: End-to-end Program Repair with LLMs & 2023 & \cite{77inferfixProgramRepair} \\
93 & Program repair & Keep the Conversation Going: Fixing 162 out of 337 bugs for \$0.42 each   using ChatGPT & 2023 & \cite{2keepConversationGoing} \\
94 & Program repair & Neural Program Repair with Program Dependence Analysis and Effective   Filter Mechanism & 2023 & \cite{71repairProgramDependence} \\
95 & Program repair & Out of Context: How important is Local Context in Neural Program Repair? & 2023 & \cite{prenner2023context} \\
96 & Program repair & Pre-trained Model-based Automated Software Vulnerability Repair: How Far   are We? & 2023 & \cite{148PreTrainedModel} \\
97 & Program repair & RAPGen: An Approach for Fixing Code Inefficiencies in Zero-Shot & 2023 & \cite{154RapgenAnApproach} \\
98 & Program repair & RAP-Gen: Retrieval-Augmented Patch Generation with CodeT5 for Automatic   Program Repair & 2023 & \cite{202RapGenRetrieval} \\
99 & Program repair & STEAM: Simulating the InTeractive BEhavior of ProgrAMmers for Automatic   Bug Fixing & 2023 & \cite{102SteamSimulatingThe} \\
100 & Program repair & Towards Generating Functionally Correct Code Edits from Natural Language   Issue Descriptions & 2023 & \cite{68codeEdits} \\
101 & Program repair & VulRepair: a T5-based Automated Software Vulnerability Repair & 2022 & \cite{43vulRepairVulnerabilityRepair} \\
102 & Program repair & What Makes Good In-Context Demonstrations for Code Intelligence Tasks   with LLMs? & 2023 & \cite{205WhatMakesGood} \\
\bottomrule
\end{tabular}
}
\end{table*}

\subsection{Paper Collection Methodology}
\label{subsec_collect_method}

\begin{figure}[t!]
\centering
\includegraphics[width=1\linewidth]{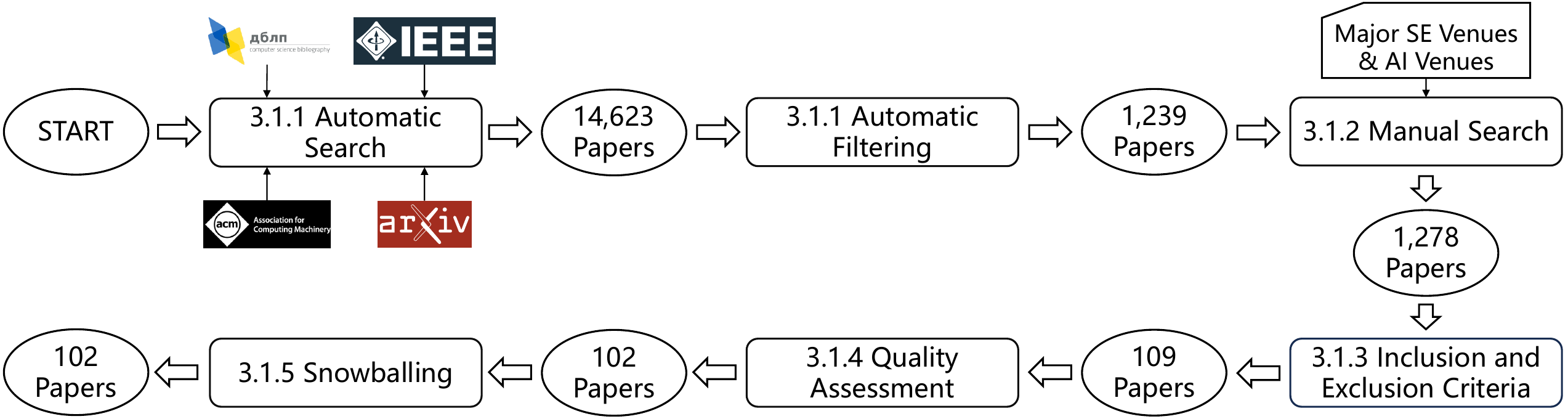}
\caption{Overview of the paper collection process}
\label{fig:paperCollection}
\vspace{-0.1in}
\end{figure}

Figure \ref{fig:paperCollection} shows our paper search and selection process. 
To collect as much relevant literature as possible, we use both automatic search (from paper repository database) and manual search (from major software engineering and artificial intelligence venues).
We searched papers from Jan. 2019 to Jun. 2023 and further conducted the second round of search to include the papers from Jul. 2023 to Oct. 2023.

\begin{table}[!t]
\centering
\caption{Conference proceedings and journals considered for manual search}
\label{tab:venues}
\scalebox{0.7}{
\begin{tabular}{p{0.2 cm}p{1.7 cm}p{9.2 cm}}
\toprule
 & \textbf{Acronym} & \textbf{Venue} \\ \midrule
\multirow{12}{*}{\rotatebox{90}{\textbf{SE Conference}}} & ICSE & International Conference on Software Engineering \\
 & ESEC/FSE & Joint European Software Engineering Conference and Symposium on the   Foundations of Software Engineering \\
 & ASE & International Conference on Automated Software Engineering \\
 & ISSTA & International Symposium on Software Testing and Analysis \\
 & ICST & International Conference on Software Testing, Verification and Validation \\
 & ESEM & International Symposium on Empirical Software Engineering and Measurement \\
 & MSR & International Conference on Mining Software Repositories \\
 & QRS & International Conference on Software Quality, Reliability and Security \\
 & ICSME & International Conference on Software Maintenance and Evolution \\
 & ISSRE & International Symposium on Software Reliability Engineering \\ \midrule
\multirow{11}{*}{\rotatebox{90}{\textbf{SE Journal}}} & TSE & Transactions on Software Engineering \\
 & TOSEM & Transactions on Software Engineering and Methodology \\
 & EMSE & Empirical Software Engineering \\
 & ASE & Automated Software Engineering \\
 & JSS & Journal of Systems and Software \\
 & JSEP & Journal of Software: Evolution and Process \\
 & STVR & Software Testing, Verification and Reliability \\
 & IEEE SOFTW. & IEEE Software \\
 & IET SOFTW. & IET Software \\
 & IST & Information and Software Technology \\
 & SQJ & Software Quality Journal \\ \midrule
\multirow{7}{*}{\rotatebox{90}{\textbf{AI Venues}}} & ICLR & International Conference on Learning Representations \\
 & NeurIPS & Conference on Neural Information Processing Systems \\
 & ICML & International Conference on Machine Learning \\
 & AAAI & AAAI Conference on Artificial Intelligence \\
 & EMNLP & Conference on Empirical Methods in Natural Language Processing \\
 & ACL & Annual Meeting of the Association for Computational Linguistics \\
 & IJCAI & International Joint Conference on Artificial Intelligence \\
\bottomrule
\end{tabular}
}
\end{table}

\subsubsection{Automatic Search}
\label{subsec_collect_method_automatic}
To ensure that we collect papers from diverse research areas, we conduct an extensive search using four popular scientific databases: ACM digital library, IEEE Xplore digital library, arXiv, and DBLP.

We search for papers whose title contains \textit{keywords related to software testing tasks and testing techniques} (as shown below) in the first three databases. 
In the case of DBLP, we use additional \textit{keywords related to LLMs} (as shown below) to filter out irrelevant studies, as relying solely on testing-related keywords would result in a large number of candidate studies.
While using two sets of keywords for DBLP may result in overlooking certain related studies, we believe it is still a feasible strategy. 
This is due to the fact that a substantial number of studies present in this database can already be found in the first three databases, and the fourth database only serves as a supplementary source for collecting additional papers.


\begin{itemize}
\item Keywords related with software testing tasks and techniques: test \textit{OR} bug \textit{OR} issue \textit{OR} defect \textit{OR} fault \textit{OR} error \textit{OR} failure \textit{OR} crash \textit{OR} debug \textit{OR} debugger \textit{OR} repair \textit{OR} fix \textit{OR} assert \textit{OR} verification \textit{OR} validation \textit{OR} fuzz \textit{OR} fuzzer \textit{OR} mutation.


\item Keywords related with LLMs: LLM \textit{OR} language model \textit{OR} generative model \textit{OR} large model \textit{OR} GPT-3 \textit{OR} ChatGPT \textit{OR} GPT-4 \textit{OR} LLaMA \textit{OR} PaLM2 \textit{OR} CodeT5 \textit{OR} CodeX \textit{OR} CodeGen \textit{OR} Bard \textit{OR} InstructGPT. 
Note that, we only list the top ten most popular LLMs (based on Google search), since they are the search keywords for matching paper titles, rather than matching the paper content.

\end{itemize}


The above search strategy based on the paper title can recall a large number of papers, and we further conduct the automatic filtering based on the paper content. 
Specifically, we filter the paper whose content contains ``LLM'' or ``language model'' or ``generative model'' or ``large model'' or the name of the LLMs (using the LLMs in~\cite{zhao2023surveyLLM,pan2023unifying} except those in our exclusion criteria).
This can help eliminate the papers that do not involve the neural models. 


\subsubsection{Manual Search}
\label{subsec_collect_method_manual}

To compensate for the potential omissions that may result from automated searches, we also conduct manual searches.  
In order to make sure we collect highly relevant papers, we conduct a manual search within the conference proceedings and journal articles from top-tier software engineering venues (listed in Table \ref{tab:venues}).

In addition, given the interdisciplinary nature of this work, we also include the conference proceedings of the artificial intelligence field. 
We select the top ten venues based on the h5 index from Google Scholar, and exclude three computer vision venues, i.e., CVPR, ICCV, ECCV, as listed in Table \ref{tab:venues}.


\subsubsection{Inclusion and Exclusion Criteria}
\label{subsec_collect_method_inclusion}

The search conducted on the databases and venue is, by design, very inclusive. 
This allows us to collect as many papers as possible in our pool. 
However, this generous inclusivity results in having papers that are not directly related to the scope of this survey. Accordingly, we define a set of specific inclusion and exclusion criteria and then we apply them to each paper in the pool and remove papers not meeting the criteria.  
This ensures that each collected paper aligns with our scope and research questions.

\textbf{Inclusion Criteria.} 
We define the following criteria for including papers: 
\begin{itemize}
    \item The paper proposes or improves an approach, study, or tool/framework that targets testing specific software or systems with LLMs. 
    \item The paper applies LLMs to software testing practice, including all tasks within the software testing lifecycle as demonstrated in Section \ref{subsec_testLifecycle}.
    \item The paper presents an empirical or experimental study about utilizing LLMs in software testing practice. 
    \item The paper involves specific testing techniques (e.g., fuzz testing) employing LLMs. 
\end{itemize}

If a paper satisfies any of the following criteria, we will include it.

\textbf{Exclusion Criteria.} 
The following studies would be excluded during study selection: 
\begin{itemize}
\item The paper does not involve software testing tasks, e.g., code comment generation.
\item The paper does not utilize LLMs, e.g., using recurrent neural networks.
\item The paper mentions LLMs only in future work or discussions rather than using LLMs in the approach.
\item The paper utilizes language models with encoder-only architecture, e.g., BERT, which can not directly be utilized for generation tasks (as demonstrated in Section \ref{subsec_background_LLM}). 
\item The paper focuses on testing the performance of LLMs, such as fairness, stability, security, etc. \cite{modelTest1, modelTest2, modelTest3}.
\item The paper focuses on evaluating the performance of LLM-enabled tools, e.g., evaluating the code quality of the code generation tool Copilot~\cite{toolTest1, toolTest2, toolTest3}.
\end{itemize}

For the papers collected through automatic search and manual search, we conduct a manual inspection to check whether they satisfy our inclusion criteria and filter those following our exclusion criteria. 
Specifically, the first two authors read each paper to carefully determine whether it should be included based on the inclusion criteria and exclusion criteria, and any paper with different decisions will be handed over to the third author to make the final decision. 


\subsubsection{Quality Assessment}
\label{subsec_collect_method_quality}

In addition, we establish quality assessment criteria to exclude low-quality studies as shown below. 
For each question, the study's quality is rated as ``yes'', ``partial'' or ``no'' which are assigned values of 1, 0.5, and 0, respectively.
Papers with a score of less than eight will be excluded from our study. 

\begin{itemize}

\item Is there a clearly stated research goal related to software testing?
\item Is there a defined and repeatable technique?
\item Is there any explicit contribution to software testing?
\item Is there an explicit description of which LLMs are utilized? 
\item Is there an explicit explanation about how the LLMs are utilized?
\item Is there a clear methodology for validating the technique? 
\item Are the subject projects selected for validation suitable for the research goals?
\item Are there control techniques or baselines to demonstrate the effectiveness of the proposed technique? 
\item Are the evaluation metrics relevant (e.g., evaluate the effectiveness of the proposed technique) to the research objectives?
\item Do the results presented in the study align with the research objectives and are they presented in a clear and relevant manner?
\end{itemize}

\subsubsection{Snowballing}
\label{subsec_collect_method_snowballing}

At the end of searching database repositories and conference proceedings and journals, and applying inclusion/exclusion criteria and quality assessment, we obtain the initial set of papers. 
Next, to mitigate the risk of omitting relevant literature from this survey, we also perform backward snowballing \cite{DBLP:conf/ease/Wohlin14} by inspecting the references cited by the collected papers so far. 
Note that, this procedure did not include new studies, which might because the surveyed topic is quite new and the reference studies tend to published previously, and we already include a relatively comprehensive automatic and manual search.



\begin{figure}[t!]
\centering
\includegraphics[width=0.8\linewidth]{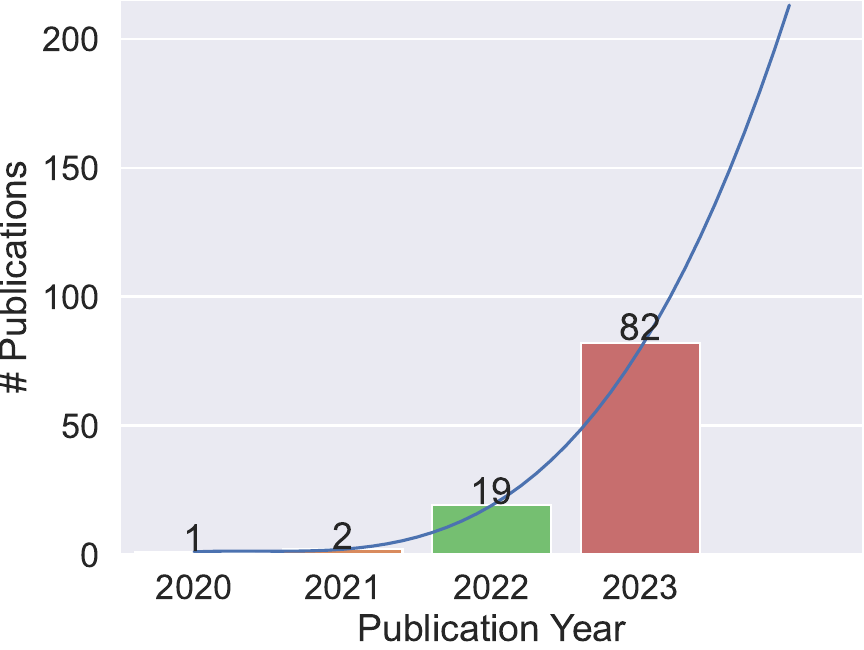}
\caption{Trend in the number of papers with year}
\label{fig:paperTrends}
\vspace{-0.1in}
\end{figure}

\subsection{Collection Results}
\label{subsec_collect_results}

As shown in Figure \ref{fig:paperCollection}, the collection process started with a total of 14,623 papers retrieved from four academic databases employing keyword searching. Then after automated filtering, manual search, applying inclusion/exclusion criteria, and quality assessment, we finally collected a total of 102 papers involving software testing with LLMs.
Table \ref{tab:paperOverview} shows the details of the collected papers.
Besides, we also use Table \ref{tab:collectedPaperDetailed} (at the end of the paper) to provide a more comprehensive overview of these papers regarding the specific characteristics which will be illustrated in Section \ref{sec_analysis_testing} and Section \ref{sec_analysis_LLM}.

Note that, there are two studies which are respectively the extension of a previously published paper by the same authors (\cite{80UsingTransferLearning} and \cite{79StudyingtheUsage}, \cite{82variablediscovery} and \cite{81largelanguagemodels}), and we only keep the extended version to avoid duplicate. 






\begin{figure*}[t!]
\centering
\includegraphics[width=\linewidth]{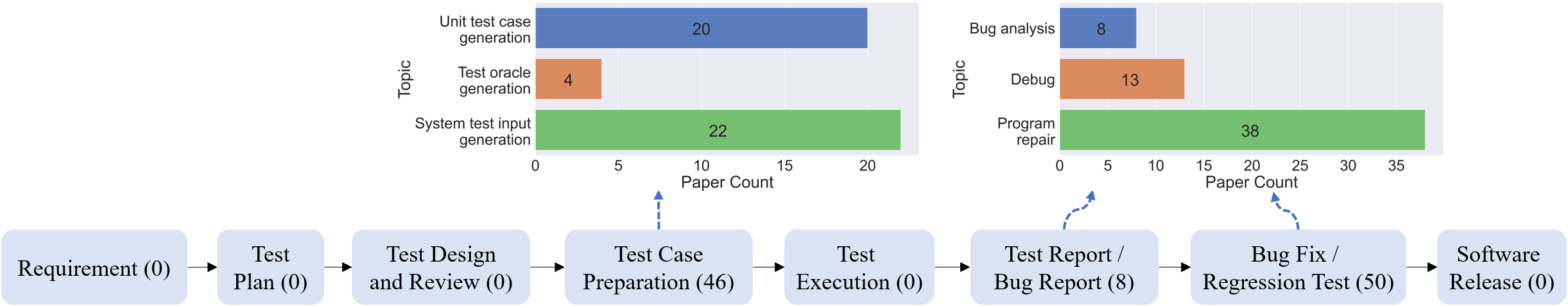}
\caption{Distribution of testing tasks with LLMs (aligned with software testing life cycle \cite{Myers2004theart,Vinay2008manage,Tchier2015Softwaretesting}, the number in bracket indicates the number of collected studies per task, and one paper might involve multiple tasks)}

\label{fig:testingTasks}
\vspace{-0.1in}
\end{figure*}

\subsection{General Overview of Collected Paper}
\label{subsec_general_overview}

Among the papers, 47\% papers are published in software engineering venues, among which 19 papers are from ICSE, 5 papers are from FSE, 5 papers are from ASE, and 3 papers are from ISSTA. 
2\% papers are published in artificial intelligence venues such as EMNLP and ICLR, and 5\% papers are published in program analysis or security venues like PLDI and S\&P. 
Besides, 46\% of the papers have not yet been published via peer-reviewed venues, i.e., they are disclosed on arXiv. 
This is understandable because this field is emerging and many works are just completed and in the process of submission. Although these papers did not undergo peer review, we have a quality assessment process that eliminates papers with low quality, which potentially ensures the quality of this survey.

Figure \ref{fig:paperTrends} demonstrates the trend of our collected papers per year. 
We can see that as the years go by, the number of papers in this field is growing almost exponentially. In 2020 and 2021, there were only 1 and 2 papers, respectively. In 2022, there were 19 papers,  
and in 2023, there have been 82 papers.
It is conceivable that there will be even more papers in the future, which indicates the popularity and attention that this field is receiving. 



\section{Analysis from Software Testing Perspective}
\label{sec_analysis_testing}

This section presents our analysis from the viewpoint of software testing and organizes the collected studies in terms of testing tasks.  
Figure \ref{fig:testingTasks} lists the distribution of each involved testing task, aligned with the software testing life cycle. 
We first provide a general overview of the distribution, followed by further analysis for each task.  
Note that, for each following subsection, the cumulative total of subcategories may not always match the total number of papers since a paper might belong to more than one subcategory.

%


We can see that LLMs have been effectively used in both the mid to late stages of the software testing lifecycle. 
In the test case preparation phase, LLMs have been utilized for tasks such as generating unit test cases, test oracle generation, and system test input generation. 
These tasks are crucial in the mid-phase of software testing to help catch issues and prevent further development until issues are resolved.
Furthermore, in later phases such as the test report/bug reports and bug fix phase, LLMs have been employed for tasks such as bug analysis, debugging, and repair. 
These tasks are critical towards the end of the testing phase when software bugs need to be resolved to prepare for the product's release.

\subsection{Unit Test Case Generation}
\label{subsec_testingtask_unit}
Unit test case generation involves writing unit test cases to check individual units/components of the software independently and ensure that they work correctly.
For a method under test (i.e., often called the focal method), its corresponding unit test consists of a test prefix and a test oracle. 
In particular, the test prefix is typically a series of method invocation statements or assignment statements, which aims at driving the focal method to a testable state; and then the test oracle serves as the specification to check whether the current behavior of the focal method satisfies the expected one, e.g., the test assertion. 

To alleviate manual efforts in writing unit tests, researchers have proposed various techniques to facilitate automated unit test generation. 
Traditional unit test generation techniques leverage search-based \cite{harman2010theoretical,pedro2023interactive}, constraint-based \cite{xiao2013characteristics} or random-based strategies \cite{pacheco2007feedback} to generate a suite of unit tests with the main goal of maximizing the coverage in the software under test.
Nevertheless, the coverage and the meaningfulness of the generated tests are still far from satisfactory.

Since LLMs have demonstrated promising results in tasks such as code generation, and given that both code generation and unit test case generation involve generating source code, recent research has extended the domain of code generation to encompass unit test case generation.
Despite initial success, there are nuances that set unit test case generation apart from general code generation, signaling the need for more tailored approaches. 


\begin{table*}[!ht]
\caption{Performance of unit test case generation}
\label{tab:unit_test}
\centering
\begin{tabular}{p{3.8cm}|p{1.5cm}|p{6cm}|p{2cm}|p{0.6cm}}
\hline
\textbf{Dataset} & \textbf{Correctness} & Coverage & LLM & Paper \\
\hline
5 Java projects from Defects4J & 16.21\% & 5\%-13\% (line coverage) & BART & \cite{28unitTest} \\
\hline
10 Jave projects & 40\% & 89\% (line coverage), 90\% (branch coverage) & ChatGPT & \cite{63chatUniTest} \\
\hline
CodeSearchNet & 41\% &  N/A & ChatGPT & \cite{64noUnitTest} \\
\hline
HumanEval &  78\% & 87\% (line coverage), 92\% (branch coverage) & Codex  & \cite{66generatingUnitTests} \\
\hline
SF110 & 2\% & 2\% (line coverage), 1\% (branch coverage)  & Codex  & \cite{66generatingUnitTests} \\
\hline
\end{tabular}
\newline Note that, \cite{66generatingUnitTests} experiments with Codex, CodeGen, and ChatGPT, and the best performance was achieved by Codex.
\vspace{-0.1in}
\end{table*}

\textbf{Pre-training or fine-tuning LLMs for unit test case generation.} 
Due to the limitations of LLMs in their earlier stages, a majority of the earlier published studies adopt this pre-training or fine-tuning schema. 
Moreover, in some recent studies, this schema continues to be employed to increase the LLMs' familiarity with domain knowledge.
Alagarsamy et al.~\cite{23a3testTestGeneration} first pre-trained the LLM with the focal method and asserted statements to enable the LLM to have a stronger foundation knowledge of assertions, then fine-tuned the LLM for the test case generation task where the objective is to learn the relationship between the focal method and the corresponding test case. 
Tufano et al. \cite{28unitTest} utilized a similar schema by pre-training the LLM on a large unsupervised Java corpus, and supervised fine-tuning a downstream translation task for generating unit tests. 
Hashtroudi et al. \cite{119AutomatedTestCase} leveraged the existing developer-written tests for each project to generate a project-specific dataset for domain adaptation when fine-tuning the LLM, which can facilitate generating human-readable unit tests.
Rao et al. \cite{111CatLmTraining} trained a GPT-style language model by utilizing a pre-training signal that explicitly considers the mapping between code and test files.
Steenhoek et al. \cite{steenhoek2023reinforcement} utilizes reinforcement learning to optimize models by providing rewards based on static quality metrics that can be automatically computed for the generated unit test cases.


\textbf{Designing effective prompts for unit test case generation.} 
The advancement of LLMs has allowed them to excel at targeted tasks without pre-training or fine-tuning. 
Therefore most later studies typically focus on how to design the prompt, to make the LLM better at understanding the context and nuances of this task. 
Xie et al. \cite{63chatUniTest} generated unit test cases by parsing the project, extracting essential information, and creating an adaptive focal context that includes a focal method and its dependencies within the pre-defined maximum prompt token limit of the LLM, and incorporating these context into a prompt to query the LLM. 
Dakhel et al. \cite{117EffectiveTestGeneration} introduced MuTAP for improving the effectiveness of test cases generated by LLMs in terms of revealing bugs by leveraging mutation testing. 
They augment prompts with surviving mutants, as those mutants highlight the limitations of test cases in detecting bugs. 
Zhang et al. \cite{112HowWellDoes} generated security tests with vulnerable dependencies with LLMs. 

Yuan et al. \cite{64noUnitTest} first performed an empirical study to evaluate ChatGPT’s capability of unit test generation with both a quantitative analysis and a user study in terms of correctness, sufficiency, readability, and usability. 
And results show that the generated tests still suffer from correctness issues, including diverse compilation errors and execution failures. 
They further propose an approach that leveraged the ChatGPT itself to improve the quality of its generated tests with an initial test generator and an iterative test refiner. 
Specifically, the iterative test refiner iteratively fixed the compilation errors in the tests generated by the initial test generator, which follows a validate-and-fix paradigm to prompt the LLM based on the compilation error messages and additional code context. 
Guilherme et al. \cite{150AnInitialInvestigation} and Li et al. \cite{113PromptingCodeInterpreter} respectively evaluated the quality of the generated unit tests by LLM using different metrics and different prompts. 



\textbf{Test generation with additional documentation.} 
Vikram et al. \cite{121CanLargeLanguage} went a step further by investigating the potential of using LLMs to generate property-based tests when provided API documentation. 
They believe that the documentation of an API method can assist the LLM in producing logic to generate random inputs for that method and deriving meaningful properties of the result to check. 
Instead of generating unit tests from the source code, Plein et al. \cite{110AutomaticGenerationOf} generated the tests based on user-written bug reports.

\textbf{LLM and search-based method for unit test generation.} 
The aforementioned studies utilize LLMs for the whole unit test case generation task, while Lemieux et al. \cite{55codamosaTestGeneration} focus on a different direction, i.e., first letting the traditional search-based software testing techniques (e.g., Pynguin \cite{Lukasczyk2022Pynguin}) in generating unit test case until its coverage improvements stall, then asking the LLM to provide the example test cases for under-covered functions. These examples can help the original test generation redirect its search to more useful areas of the search space. 

Tang et al. \cite{123ChatgptVsSbst} conducts a systematic comparison of test suites generated by the LLM and the state-of-the-art search-based software testing tool EvoSuite, by considering the correctness, readability, code coverage, and bug detection capability.
Similarly, Bhatia \cite{bhatia2023unit} experimentally investigates the quality of unit tests generated by LLM compared to a commonly-used test generator Pynguin.

\textbf{Performance of unit test case generation.} 
Since the aforementioned studies of unit test case generation are based on different datasets, one can hardly derive a fair comparison and we present the details in Table \ref{tab:unit_test} to let the readers obtain a general view.  
We can see that in the SF110 benchmark, all three evaluated LLMs have quite low performance, i.e., 2\% coverage \cite{66generatingUnitTests}. 
SF110 is an Evosuite (a search-based unit test case generation technique) benchmark consisting of 111 open-source Java projects retrieved from SourceForge, containing 23,886 classes, over 800,000 bytecode-level branches, and 6.6 million lines of code.
The authors did not present detailed reasons for the low performance which can be further explored in the future.

\subsection{Test Oracle Generation} 
\label{subsec_testingtask_oracle}
A test oracle is a source of information about whether the output of a software system (or program or function or method) is correct or not~\cite{barr2014oracle}.  
Most of the collected studies in this category target the test assertion generation, which is inside a unit test case. 
Nevertheless, we opted to treat these studies as separate sections to facilitate a more thorough analysis. 


Test assertion, which is to indicate the potential issues in the tested code, is an important aspect that can distinguish the unit test cases from the regular code. 
This is why some studies specifically focus on the generation of effective test assertions. 
Actually, before using LLMs, researchers have proposed RNN-based approaches that aim at learning from thousands of unit test methods to generate meaningful assert statements \cite{cody2020onlearning}, yet only 17\% of the generated asserts can exactly match with the ground truth asserts.
Subsequently, to improve the performance, several researchers utilized the LLMs for this task. 

Mastropaolo et al.\cite{79StudyingtheUsage, 80UsingTransferLearning} pre-trained a T5 model on a dataset composed of natural language English text and source code. 
Then, it fine-tuned such a model by reusing datasets used in four previous works that used deep learning techniques (such as RNN as mentioned before) including test assertion generation and program repair, etc. 
Results showed that the extract match rate of the generated test assertion is 57\%. 
Tufano et al. \cite{39assertStatements} proposed a similar approach which separately pre-trained the LLM with English corpus and code corpus, and then fine-tuned it on the asserts dataset (with test methods, focal methods, and asserts).
This further improved the performance to 62\% of the exact match rate. 
Besides the syntax-level data as previous studies, Nie et al. \cite{24semanticsTestCompletion} fine-tuned the LLMs with six kinds of code semantics data, including the execution result (e.g., types of the local variables) and execution context (e.g., the last called method in the test method), which enabled LLMs to learn to understand the code execution information. 
The exact match rate is 17\% (note that this paper is based on a different dataset from all other studies mentioned under this topic).

The aforementioned studies utilized the pre-training and fine-tuning schema when using LLMs, and with the increasingly powerful capabilities of LLMs, they can perform well on specific tasks without these specialized pre-training or fine-tuning datasets. 
Subsequently, Nashid et al. \cite{56retrievalPromptSelection} utilized prompt engineering for this task, and proposed a technique for prompt creation that automatically retrieves code demonstrations similar to the task, based on embedding or frequency analysis.
They also present evaluations about the few-shot learning with various numbers (e.g., zero-shot, one-shot, or n-shot) and forms (e.g., random vs. systematic, or with vs. without natural language descriptions) of the prompts, to investigate its feasibility on test assertion generation. 
With only a few relevant code demonstrations, this approach can achieve an accuracy of 76\% for exact matches in test assertion generation, which is the state-of-the-art performance for this task. 


\subsection{System Test Input Generation}
\label{subsec_testingtask_input}


This category encompasses the studies related to creating test input of system testing for enabling the automation of test execution.
We employ three subsections to present the analysis from three different orthogonal viewpoints, and each of the collected studies may be analyzed in one or more of these subsections.

The first subsection is \textit{input generation in terms of software types}.
The generation of system-level test inputs for software testing varies for specific types of software being tested.
For example, for mobile applications, the test input generation requires providing a diverse range of text inputs or operation combinations (e.g., click a button, long press a list) \cite{26fillBlank,60GUITesting}, which is the key to testing the application's functionality and user interface; while for Deep Learning (DL) libraries, the test input is a program which covers diversified DL APIs \cite{13fuzzDeepLearningLibraries,19fuzzDeepLearningLibraries}. 
This subsection will demonstrate how the LLMs are utilized to generate inputs for different types of software.

The second subsection \textit{input generation in terms of testing techniques}.
We have observed that certain approaches serve as specific types of testing techniques. 
For example, dozens of our collected studies specifically focus on using LLMs for fuzz testing. 
Therefore, this subsection would provide an analysis of the collected studies in terms of testing techniques, showcasing how the LLMs are employed to enhance traditional testing techniques.

The third subsection \textit{input generation in terms of input and output}.
While most of the collected studies take the source code or the software itself as the input and directly output the software's test input, there are studies that utilize alternative forms of input and output. 
This subsection would provide an analysis of such studies, highlighting different approaches and their input-output characteristics.



\subsubsection{Input Generation in Terms of Software Types}
\label{subsec_systemInput_softwareType}

\begin{figure}[!ht]
\centering
\includegraphics[width=\linewidth]{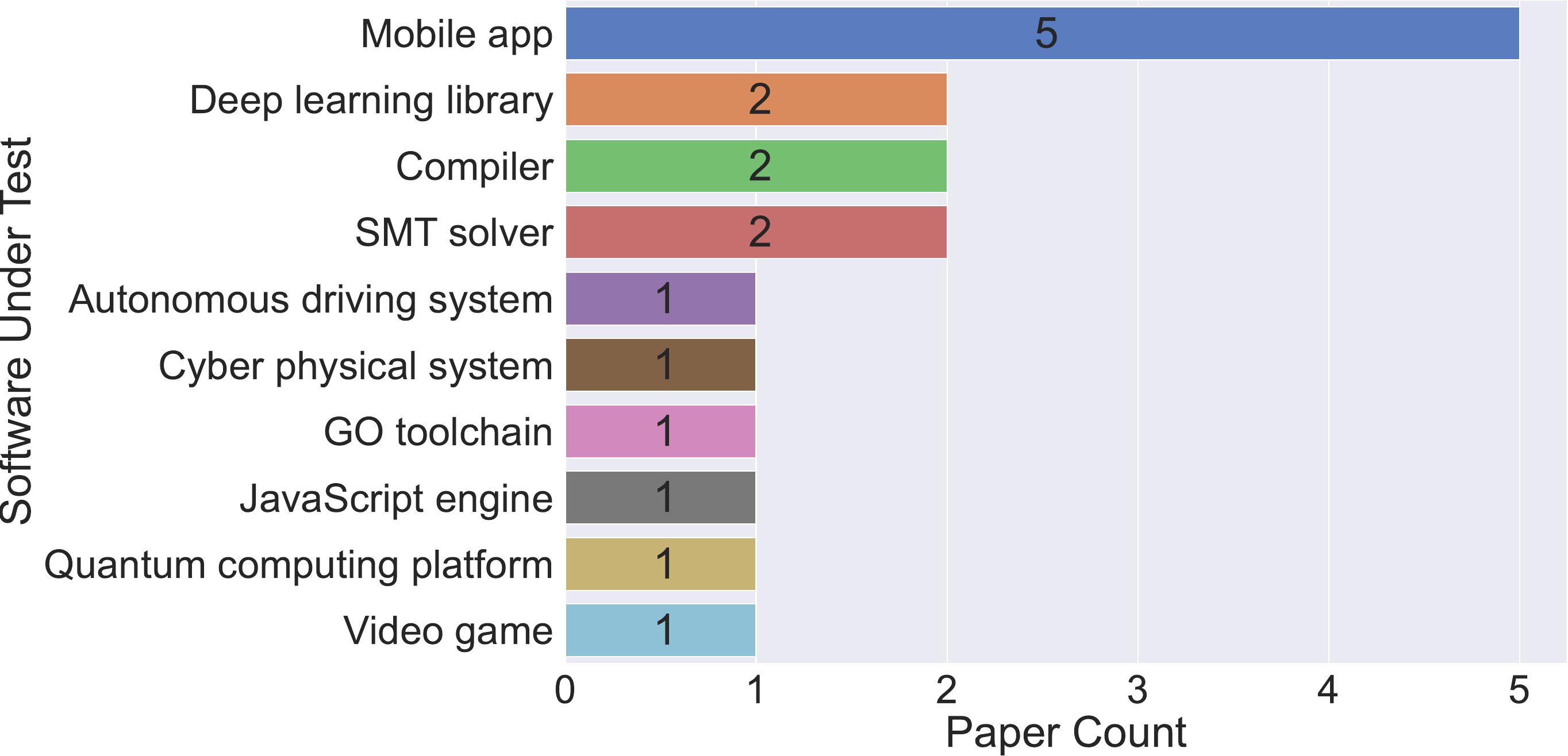}
\caption{Distribution of software under test}
\label{fig:softwareUnderTest}
\vspace{-0.1in}
\end{figure}

Figure \ref{fig:softwareUnderTest} demonstrates the types of software under test in our collected studies. 
It is evident that the most prominent category is mobile apps, with five studies utilizing LLMs for testing, possibly due to their prevalence and importance in today's business and daily life.
Additionally, there are respectively two studies focusing on testing deep learning libraries, compilers, and SMT solvers.
Moreover, LLM-based testing techniques have also been applied to domains such as cyber-physical systems, quantum computing platforms, and more.
This widespread adoption of LLMs demonstrates their effectiveness in handling diverse test inputs and enhancing testing activities across various software domains. 
A detailed analysis is provided below.

\textbf{Test input generation for mobile apps.}
For mobile app testing, one difficulty is to generate the appropriate text inputs to proceed to the next page, which remains a prominent obstacle for testing coverage. 
Considering the diversity and semantic requirement of valid inputs (e.g., flight departure, movie name), traditional techniques with heuristic-based or constraint-based techniques \cite{li2017droidbot,he2020textexerciser} are far from generating meaningful text input. 
Liu et al. \cite{26fillBlank} employ the LLM to intelligently generate the semantic input text according to the GUI context.
In detail, their proposed QTypist automatically extracts the component information related to the EditText for generating the prompts, and then inputs the prompts into the LLM to generate the input text.

Besides the text input, there are other forms of input for mobile apps, i.e., operations like `click a button' and `select a list'.
To fully test an app, it is required to cover more GUI pages and conduct more meaningful exploration traces through the GUI operations, yet existing studies with random-/rule-based methods \cite{Monkey,li2017droidbot}, model-based methods \cite{su2017guided,dong2020time}, and learning-based methods \cite{pan2020reinforcement} are unable to understand the semantic information of the GUI page thus could not conduct the trace planning effectively. 
Liu et al. \cite{60GUITesting} formulates the test input generation of mobile GUI testing problem as a Q\&A task, which asks LLM to chat with the mobile apps by passing the GUI page information to LLM to elicit testing scripts (i.e., GUI operation), and executing them to keep passing the app feedback to LLM, iterating the whole process. 
The proposed GPTDroid extracts the static context of the GUI page and the dynamic context of the iterative testing process, and designs prompts for inputting this information to LLM which enables the LLM to better understand the GUI page as well as the whole testing process. 
It also introduces a functionality-aware memory prompting mechanism that equips the LLM with the ability to retain testing knowledge of the whole process and conduct long-term, functionality-based reasoning to guide exploration.
Similarly, Zimmermann et al. utilize the LLM to interpret natural language test cases and programmatically navigate through the application under test \cite{84automatingGUI}.

Yu et al. \cite{115LlmForTest} investigate the LLM's capabilities in the mobile app test script generation and migration task, including the scenario-based test generation, and the cross-platform/app test migration.

\textbf{Test input generation for DL libraries.}
The input for testing DL libraries is DL programs, and the difficulty in generating the diversified input DL programs is that they need to satisfy both the input language (e.g., Python) syntax/semantics and the API input/shape constraints for tensor computations.
Traditional techniques with API-level fuzzing \cite{wei2022freelunch,xie2022docter} or model-level fuzzing \cite{guo2020audee,wang2020deep} suffer from the following limitations: 1) lack of diverse API sequence thus cannot reveal bugs caused by chained API sequences; 2) cannot generate arbitrary code thus cannot explore the huge search space that exists when using the DL libraries. 
Since LLMs can include numerous code snippets invoking DL library APIs in their training corpora, they can implicitly learn both language syntax/semantics and intricate API constraints for valid DL program generation. 
Taken in this sense, Deng et al. \cite{13fuzzDeepLearningLibraries} used both generative and infilling LLMs to generate and mutate valid/diverse input DL programs for fuzzing DL libraries.
In detail, it first uses a generative LLM (CodeX) to generate a set of seed programs (i.e., code snippets that use the target DL APIs). 
Then it replaces part of the seed program with masked tokens using different mutation operators and leverages the ability of infilling LLM (InCoder) to perform code infilling to generate new code that replaces the masked tokens. 
Their follow-up study \cite{19fuzzDeepLearningLibraries} goes a step further to prime LLMs to synthesize unusual programs for the fuzzing DL libraries. 
It is built on the well-known hypothesis that historical bug-triggering programs may include rare/valuable code ingredients important for bug finding and show improved bug detection performance.



\textbf{Test input generation for other types of software.} 
There are also dozens of studies that address testing tasks in various other domains, due to space limitations, we will present a selection of representative studies in these domains.

Finding bugs in a commercial cyber-physical system (CPS) development tool such as Simulink is even more challenging. 
Given the complexity of the Simulink language, generating valid Simulink model files for testing is an ambitious task for traditional machine learning or deep learning techniques. 
Shrestha et al. \cite{12slgptSimulink} employs a small set of Simulink-specific training data to fine-tune the LLM for generating Simulink models.
Results show that it can create Simulink models quite similar to the open-source models, and can find a super-set of the bugs traditional fuzzing approaches found. 

Sun et al. \cite{204SmtSolverValidation} utilize LLM to generate test formulas for fuzzing SMT solvers.
It retrains the LLMs on a large corpus of SMT formulas to enable them to acquire SMT-specific domain knowledge. 
Then it further fine-tunes the LLMs on historical bug-triggering formulas, which are known to involve structures that are more likely to trigger bugs and solver-specific behaviors. 
The LLM-based compiler fuzzer proposed by Yang et al. \cite{175WhiteBoxCompiler} adopts a dual-model framework: (1) an analysis LLM examines the low-level optimization source code and produces requirements on the high-level test programs that can trigger the optimization; (2) a generation LLM produces test programs based on the summarized requirements.
Ye et al. \cite{14conformanceTesting} utilize the LLM for generating the JavaScript programs and then use the well-structured ECMAScript specifications to automatically generate test data along with the test programs, after that they apply differential testing to expose bugs. 


\subsubsection{Input Generation in Terms of Testing Techniques}
\label{subsec_systemInput_testTechnique}




By utilizing system test inputs generated by LLMs, the collected studies aim to enhance traditional testing techniques and make them more effective. 
Among these techniques, fuzz testing is the most commonly involved one.
Fuzz testing, as a general concept, revolves around generating invalid, unexpected, or random data as inputs to evaluate the behavior of software. 
LLMs play a crucial role in improving traditional fuzz testing by facilitating the generation of diverse and realistic input data. 
This enables fuzz testing to uncover potential bugs in the software by subjecting it to a wide range of input scenarios.
In addition to fuzz testing, LLMs also contribute to enhancing other testing techniques, which will be discussed in detail later.

\textbf{Universal fuzzing framework.} 
Xia et al. \cite{166Fuzz4allUniversalFuzzing} present Fuzz4All that can target many different input languages and many different features of these languages. 
The key idea behind it is to leverage LLMs as an input generation and mutation engine, which enables the approach to produce diverse and realistic inputs for any practically relevant language.
To realize this potential, they present a novel auto-prompting technique, which creates LLM prompts that are well-suited for fuzzing, and a novel LLM-powered fuzzing loop, which iteratively updates the prompt to create new fuzzing inputs. 
They experiment with six different languages (C, C++, Go, SMT2, Java and Python) as inputs and demonstrate higher coverage than existing language-specific fuzzers. 
Hu et al. \cite{141AugmentingGreyboxFuzzing} propose a greybox fuzzer augmented by the LLM, which picks a seed in the fuzzer’s seed pool and prompts the LLM to produce the mutated seeds that might trigger a new code region of the software.
They experiment with three categories of input formats, i.e., formatted data files (e.g., json, xml), source code in different programming languages (e.g., JS, SQL, C), text with no explicit syntax rules (e.g., HTTP response, md5 checksum).
In addition, effective fuzzing relies on the effective fuzz driver, and  Zhang et al. \cite{140UnderstandingLargeLanguage} utilize LLMs on the fuzz driver generation, in which five query strategies are designed and analyzed from basic to enhanced. 

\textbf{Fuzzing techniques for specific software.}
There are studies that focus on the fuzzing techniques tailored to specific software, e.g., the deep learning library\cite{13fuzzDeepLearningLibraries,19fuzzDeepLearningLibraries}, compiler \cite{175WhiteBoxCompiler}, SMT solvers \cite{204SmtSolverValidation}, input widget of mobile app \cite{107TestingTheLimits}, cyber-physical system \cite{12slgptSimulink}, etc. 
One key focus of these fuzzing techniques is to generate diverse test inputs so as to achieve higher coverage. 
This is commonly achieved by combining the mutation technique with LLM-based generation, where the former produces various candidates while the latter is responsible for generating the executable test inputs \cite{13fuzzDeepLearningLibraries,204SmtSolverValidation}. 
Another focus of these fuzzing techniques is to generate the risky test inputs that can trigger bugs earlier. 
To achieve this, a common practice is to collect the historical bug-triggering programs to fine-tune the LLM \cite{204SmtSolverValidation} or treat them as the demonstrations when querying the LLM \cite{19fuzzDeepLearningLibraries,107TestingTheLimits}.

\textbf{Other testing techniques.}
There are studies that utilize LLMs for enhancing GUI testing for generating meaningful text input \cite{26fillBlank} and functionality-oriented exploration traces \cite{60GUITesting}, which has been introduced in \textit{Test input generation for mobile apps} part of Section \ref{subsec_systemInput_softwareType}.

Besides, Deng et al. \cite{120PentestgptAnLlm} leverage the LLMs to carry out penetration testing tasks automatically.
It involves setting a penetration testing goal for the LLM, soliciting it for the appropriate operation to execute, implementing it in the testing environment, and feeding the test outputs back to the LLM for next-step reasoning. 

\subsubsection{Input Generation in Terms of Input and Output}
\label{subsec_systemInput_inputOutput}


\textbf{Other output format of test generation.}
Although most works use LLM to generate test cases directly, there are also some works generating indirect inputs like testing code, test scenarios, metamorphic relations, etc.
Liu et al. \cite{107TestingTheLimits} propose InputBlaster which leverages the LLM to automatically generate unusual text inputs for fuzzing the text input widgets in mobile apps.
It formulates the unusual inputs generation problem as a task of producing a set of test generators, each of which can yield a batch of unusual text inputs under the same mutation rule. 
In detail, InputBlaster leverages LLM to produce the test generators together with the mutation rules serving as the reasoning chain and utilizes the in-context learning schema to demonstrate the LLM with examples for boosting the performance.
Deng et al. \cite{62targetTestGeneration} use LLM to extract key information related to the test scenario from a traffic rule, and represent the extracted information in a test scenario schema, then synthesize the corresponding scenario scripts to construct the test scenario.
Luu et al. \cite{106CanChatgptAdvance} examine the effectiveness of LLM in generating metamorphic relations (MRs) for metamorphic testing. 
Their results show that ChatGPT can be used to advance software testing intelligence by proposing MRs candidates that can be later adapted for implementing tests, but human intelligence should still inevitably be involved to justify and rectify their correctness.


\textbf{Other input format of test generation. }
The aforementioned studies primarily take the source code or the software as the input of LLM, yet there are also studies that take natural language description as the input for test generation.
Mathur et al. \cite{143AutomatedTestCase} propose to generate test cases from the natural language described requirements.  
Ackerman et al. \cite{152LargeLanguageModels} generate the instances from natural language described requirements recursively to serve as the seed examples for a mutation fuzzer. 

\subsection{Bug Analysis} 
This category involves analyzing and categorizing the identified software bugs to enhance understanding of the bug, and facilitate subsequent debug and bug repair.
Mukherjee et al. \cite{134EmployingDeepLearning} generate relevant answers to follow-up questions for deficient bug reports to facilitate bug triage.
Su et al. \cite{145StillConfusingFor} transform the bug-component triaging into a multi-classification task and a generation task with LLM, then ensemble the prediction results from them to improve the performance of bug-component triaging further.
Zhang et al. \cite{132CupidLeveragingChatgpt} first leverage
the LLM under the zero-shot setting to get essential information on bug reports, then use the essential information as the input to detect duplicate bug reports. 
Mahbub et al. \cite{5explainingSoftwareBugs} proposes to explain software bugs with LLM, which generates natural language explanations for software bugs by learning from a large corpus of bug-fix commits.
Zhang et al. \cite{1itigerIssueTitle} target to automatically generate the bug title from the descriptions of the bug, which aims to help developers write issue titles and facilitate the bug triaging and follow-up fixing process.

\subsection{Debug}

This category refers to the process of identifying and locating the cause of a software problem (i.e., bug). It involves analyzing the code, tracing the execution flow, collecting error information to understand the root cause of the issue, and fixing the issue.
Some studies concentrate on the comprehensive debug process, while others delve into specific sub-activities within the process.

\textbf{Overall debug framework.}
Bui et al. \cite{31detectLocalizeRepair} proposes a unified Detect-Localize-Repair framework based on the LLM for debugging, which first determines whether a given code snippet is buggy or not, then identifies the buggy lines, and translates the buggy code to its fixed version.  
Kang et al. \cite{54explainableDebugging} proposes automated scientific debugging, a technique that given buggy code and a bug-revealing test, prompts LLMs to automatically generate hypotheses, uses debuggers to actively interact with buggy code, and thus automatically reaches conclusions prior to patch generation.
Chen et al. \cite{139TeachingLargeLanguage} demonstrate that self-debugging can teach the LLM to perform rubber duck debugging; i.e., without any human feedback on the code correctness or error messages, the model is able to identify its mistakes by investigating the execution results and explaining the generated code in natural language. 
Cao et al. \cite{52promptDesign} conducts a study of LLM's debugging ability for deep learning programs, including fault detection, fault localization and program repair. 

\textbf{Bug localization.}
Wu et al. \cite{136LargeLanguageModels} compare the two LLMs (ChatGPT and GPT-4) with the existing fault localization techniques, and investigate the consistency of LLMs in fault localization, as well as how prompt engineering and the length of code context affect the results. 
Kang et al. \cite{163APreliminaryEvaluation} propose AutoFL, an automated fault localization technique that only requires a single failing test, and during its fault localization process, it also generates an explanation about why the given test fails.
Yang et al. \cite{168LargeLanguageModels} propose LLMAO to overcome the left-to-right nature of LLMs by fine-tuning a small set of bidirectional
adapter layers on top of the representations learned by LLMs, which can locate buggy lines of code without any test coverage information.
Tu et al. \cite{122Llm4cbiTamingLlms} propose LLM4CBI to tame LLMs to generate effective test programs for finding suspicious files.

\textbf{Bug reproduction.}
There are also studies focusing on a sub-phase of the debugging process. 
For example, Kang et al. \cite{9generalBugReproduction} and Plein et al. \cite{130CanLlmsDemystify} respectively propose the framework to harness the LLM to reproduce bugs, and suggest bug reproducing test cases to the developer for facilitating debugging. 
Li et al. \cite{67failureTestCases} focus on a similar aspect of finding the failure-inducing test cases whose test input can trigger the software's fault. 
It synergistically combines LLM and differential testing to do that.

There are also studies focusing on the bug reproduction of mobile apps to produce the replay script. 
Feng et al. \cite{161PromptingIsAll} propose AdbGPT, a new lightweight approach to automatically reproduce the bugs from bug reports through prompt engineering, without any training and hard-coding effort. 
It leverages few-shot learning and chain-of-thought reasoning to elicit human knowledge and logical reasoning from LLMs to accomplish the bug replay in a manner similar to a developer. 
Huang et al. \cite{173CrashtranslatorAutomaticallyReproducing} propose CrashTranslator to automatically reproduce bugs directly from the stack trace. 
It accomplishes this by leveraging the LLM to predict the exploration steps for triggering the crash, and designing a reinforcement learning based technique to mitigate the inaccurate prediction and guide the search holistically.
Taeb et al. \cite{159AxnavReplayingAccessibility} convert the manual accessibility test instructions into replayable, navigable videos by using LLM and UI element detection models, which can also help reveal accessibility issues.

\textbf{Error explanation.}
Taylor et al. \cite{164DccHelpGenerating} integrates the LLM into the Debugging C Compiler to generate unique, novice-focused explanations tailored to each error. 
Widjojo et al. \cite{171AddressingCompilerErrors} study the effectiveness of Stack Overflow and LLMs at explaining compiler errors.

\subsection{Program Repair}
\label{subsec_testingtask_repair}

This category denotes the task of fixing the identified software bugs.
The high frequency of repair-related studies can be attributed to the close relationship between this task and the source code. 
With their advanced natural language processing and understanding capabilities, LLM are well-equipped to process and analyze source code, making them an ideal tool for performing code-related tasks such as fixing bugs.

There have been template-based \cite{jiang2018shaping}, heuristic-based \cite{wen2018contextaware}, and constraint-based \cite{xiong2017precise,xuan2017nopol} automatic program repair techniques.
And with the development of deep learning techniques in the past few years, there have been several studies employing deep learning techniques for program repair. 
They typically adopt deep learning models to take a buggy software program as input and generate a patched program. 
Based on the training data, they would build a neural network model that learns the relations between the buggy code and the corresponding fixed code.
Nevertheless, these techniques still fail to fix a large portion of bugs, and they typically have to generate hundreds to thousands of candidate patches and take hours to validate these patches to fix enough bugs. 
Furthermore, the deep learning based program repair models need to be trained with huge amounts of labeled training data (typically pairs of buggy and fixed code), which is time- and effort-consuming to collect the high-quality dataset. 
Subsequently, with the popularity and demonstrated capability of the LLMs, researchers begin to explore the LLMs for program repair. 


\textbf{Patch single-line bugs.}
In the early era of program repair, the focus was mainly on addressing defects related to single-line code errors, which are relatively simple and did not require the repair of complex program logic. 
Lajk{\'o} et al. \cite{38javaScriptProgram} propose to fine-tune the LLM with JavaScript code snippets to serve as the purpose for the JavaScript program repair.
Zhang et al. \cite{71repairProgramDependence} employs program slicing to extract contextual information directly related to the given buggy statement as repair ingredients from the corresponding program dependence graph, which makes the fine-tuning more focused on the buggy code.
Zhang et al. \cite{102SteamSimulatingThe} propose a stage-wise framework STEAM for patching single-line bugs, which simulates the interactive behavior of multiple programmers involved in bug management, e.g., bug reporting, bug diagnosis, patch generation, and patch verification.

Since most real-world bugs would involve multiple lines of code, and later studies explore these more complex situations (although some of them can also patch the single-line bugs).

\textbf{Patch multiple-lines bugs.}
The studies in this category would input a buggy function to the LLM, and the goal is to output the patched function, which might involve complex semantic understanding, code hunk modification, as well as program refactoring.  
Earlier studies typically employ the fine-tuning strategy to enable the LLM to better understand the code semantics. 
Fu et al. \cite{43vulRepairVulnerabilityRepair} fine-tune the LLM by employing BPE tokenization to handle Out-Of-Vocabulary (OOV) issues which makes the approach generate new tokens that never appear in a training function but are newly introduced in the repair.
Wang et. al. \cite{202RapGenRetrieval} train the LLM  based on both buggy input and retrieved bug-fix examples which are retrieved in terms of the lexical and semantical similarities. 
The aforementioned studies (including the ones in patching single-line bugs) would predict the fixed programs directly, and Hu et al. \cite{10neuralSymbolic} utilize a different setup that predicts the scripts that can fix the bugs when executed with the delete and insert grammar. 
For example, it predicts whether an original line of code should be deleted, and what content should be inserted. 

\begin{table*}[!ht]
\caption{Performance of program repair}
\label{tab:repair}
\centering
\begin{tabular}{p{3.2cm}|p{8cm}|p{4cm}|p{0.8cm}}
\hline
\textbf{Dataset} & \textbf{\% Correct patches} & \textbf{LLM} & \textbf{Paper} \\
\hline
Defects4J v1.2, Defects4J v2.0, QuixBugs, HumanEval-Java & 22/40 Jave bugs (QuixBugs dataset, with InCoder-6B, correct code infilling setting) & PLBART, CodeT5, CodeGen, InCoder (each with variant parameters, 10 LLMs in total) & \cite{32impactLanguageModels} \\ 
\hline
QuixBugs & 23/40 Python bugs, 14/40 Java bugs (complete function generation setting) & Codex-12B & \cite{44codexQuixBugs} \\
\hline 
Defects4J v1.2, Defects4J v2.0, QuixBugs, ManyBugs & 39/40 Python bugs, 34/40 Java bugs (QuixBugs dataset, with Codex-12B, correct code infilling setting); 37/40 Python bugs, 32/40 Java bugs (QuixBugs dataset, with Codex-12B, complete function generation setting) 
& Codex, GPT-Neo, CodeT5, InCoder (each with variant parameters, 9 LLMs in total) & \cite{33practicalProgramRepair} \\
\hline
QuixBugs & 31/40 Python bugs (completion function generation setting) & ChatGPT-175B & \cite{4analysisBugFixing} \\
\hline
DL programs from StackOverflow & 16/72 Python bugs (complete function generation setting) & ChatGPT-175B & \cite{52promptDesign} \\
\hline
\end{tabular}
\newline Note that, for studies with multiple datasets or LLMs, we only present the best performance or in the most commonly utilized dataset.
\vspace{-0.1in}
\end{table*}

Nevertheless, fine-tuning may face limitations in terms of its reliance on abundant high-quality labeled data, significant computational resources, and the possibility of overfitting. 
To approach the program repair problem more effectively, later studies focus on how to design an effective prompt for program repair.
Several studies empirically investigate the effectiveness of prompt variants of the latest LLMs for program repair under different repair settings and commonly-used benchmarks (which will be explored in depth later),  
while other studies focus on proposing new techniques.
Ribeiro et al. \cite{41repairAsCompletion} take advantage of LLM to conduct the code completion in a buggy line for patch generation, and elaborate on how to circumvent the open-ended nature of code generation to appropriately fit the new code in the original program.
Xia et al. \cite{2keepConversationGoing} propose the conversation-driven program repair approach
that interleaves patch generation with instant feedback to perform the repair in a conversational style. 
They first feed the LLM with relevant test failure information to start with, and then learns from both failures and successes of earlier patching attempts of the same bug for more powerful repair. 
For earlier patches that failed to pass all tests, they combine the incorrect patches with their corresponding relevant test failure information to construct a new prompt for the
LLM to generate the next patch, in order to avoid making the same mistakes. 
For earlier patches that passed all the tests (i.e., plausible patches), they further ask the LLM to generate alternative variations of the original plausible patches.
This can further build on and learn from earlier successes to generate more
plausible patches to increase the chance of having correct patches.
Zhang et al. \cite{8bugsPythonAssignments} propose a similar approach design by leveraging multimodal prompts (e.g., natural language description, error message, input-output-based test cases), iterative querying, test-case-based few-shot selection to produce repairs.
Moon et al. \cite{moon2023coffee} propose for bug fixing with feedback. It consists of a critic model to generate feedback, an editor to edit codes based on the feedback, and a feedback selector to choose the best possible feedback from the critic.

Wei et. al. \cite{201CopilotingTheCopilots} propose Repilot to copilot the AI ``copilots'' (i.e., LLMs) by synthesizing more valid patches during the repair process. 
Its key insight is that many LLMs produce outputs autoregressively (i.e., token by token), and by resembling human writing programs, the repair can be significantly boosted and guided through a completion engine.
Brownlee et al. \cite{167EnhancingGeneticImprovement} propose to use the LLM as mutation operators for the search-based techniques of program repair.

\textbf{Repair with static code analyzer.}
Most of the program repair studies would suppose the bug has been detected, while Jin et al. \cite{77inferfixProgramRepair} propose a program repair framework paired with a static analyzer to first detect the bugs, and then fix them. 
In detail, the static analyzer first detects an error (e.g., null pointer dereference) and the context information provided by the static analyzer will be sent into the LLM for querying the patch for this specific error. 
Wadhwa et al. \cite{135FrustratedWithCode} focus on a similar task, and additionally employ an LLM as the ranker to assess the likelihood of acceptance of generated patches which can effectively catch plausible
but incorrect fixes and reduce developer burden.

\textbf{Repair for specific bugs.}
The aforementioned studies all consider the buggy code as the input for the automatic program repair, while other studies conduct program repairing in terms of other types of bug descriptions, specific types of bugs, etc.
Fakhoury et al. \cite{68codeEdits} focus on program repair from natural language issue descriptions, i.e., generating the patch with the bug and fix-related information described in the issue reports. 
Garg et al. \cite{154RapgenAnApproach} aim at repairing performance issues, in which they first retrieve a prompt instruction from a pre-constructed knowledge-base of previous performance bug fixes and then generate a repair prompt using the retrieved instruction.
There are studies focusing on the bug fixing of Rust programs \cite{104FixingRustCompilation} or OCaml programs (an industrial-strength programming language) \cite{151Gpt3Powered}.

\textbf{Empirical study about program repair.}
There are several studies related to the empirical or experimental evaluation of the various LLMs on program repair, and we summarize the performance in Table \ref{tab:repair}.
Jiang et al. \cite{32impactLanguageModels}, Xia et al. \cite{33practicalProgramRepair}, and Zhang et. al. \cite{148PreTrainedModel} respectively conduct comprehensive experimental evaluations with various LLMs and on different automated program repair benchmarks, while other researchers ~\cite{44codexQuixBugs,4analysisBugFixing,52promptDesign,144AnEvaluationOf} focus on a specific LLM and on one dataset, e.g., QuixBugs.
In addition, Gao et al. \cite{205WhatMakesGood} empirically investigate the impact of in-context demonstrations for bug fixing, including the selection, order, and number of demonstration examples. 
Prenner et al. \cite{prenner2023context} empirically study how the local context (i.e., code that comes before or after the bug location) affects the repair performance. 
Horv{\'{a}}th et al. \cite{147AnExtensiveStudy} empirically study the impact of program representation and model architecture on the repair performance.

There are two commonly-used repair settings when using LLMs to generate patches: 1) complete function generation (i.e., generating the entire patch function), 2) correct code infilling (i.e., filling in a chunk of code given the prefix and suffix), and different studies might utilize different settings which are marked in Table \ref{tab:repair}. 
The commonly-used datasets are QuixBugs, Defects4J, etc. 
These datasets only involve the fundamental functionalities such as sorting algorithms, each program’s average number of lines ranging from 13 to 22, implementing one functionality, and involving few dependencies. 
To tackle this, Cao et al. \cite{52promptDesign} conducts an empirical study on a more complex dataset with DL programs collected from StackOverflow. 
Every program contains about 46 lines of code on average, implementing several functionalities including data preprocessing, DL model construction, model training, and evaluation. 
And the dataset involves more than 6 dependencies for each program, including TensorFlow, Keras, and Pytorch. 
Their results demonstrate a much lower rate of correct patches than in other datasets, which again reveals the potential difficulty of this task.
Similarly, Haque et al. \cite{149FixevalExecutionBased} introduce a dataset comprising of buggy code submissions and their corresponding fixes collected from online judge platforms, in which it offers an extensive collection of unit tests to enable the evaluations about the correctness of fixes and further information regarding time, memory constraints, and acceptance based on a verdict.

\section{Analysis from LLM Perspective}
\label{sec_analysis_LLM}

This section discusses the analysis based on the viewpoints of LLM, specifically, it's unfolded from the viewpoints of utilized LLMs, types of prompt engineering, input of the LLMs, as well as the accompanied techniques when utilizing LLM.

\subsection{LLM Models}
\label{subsec_LLM Models}

\begin{figure}[t!]
\centering
\includegraphics[width=0.9\linewidth]{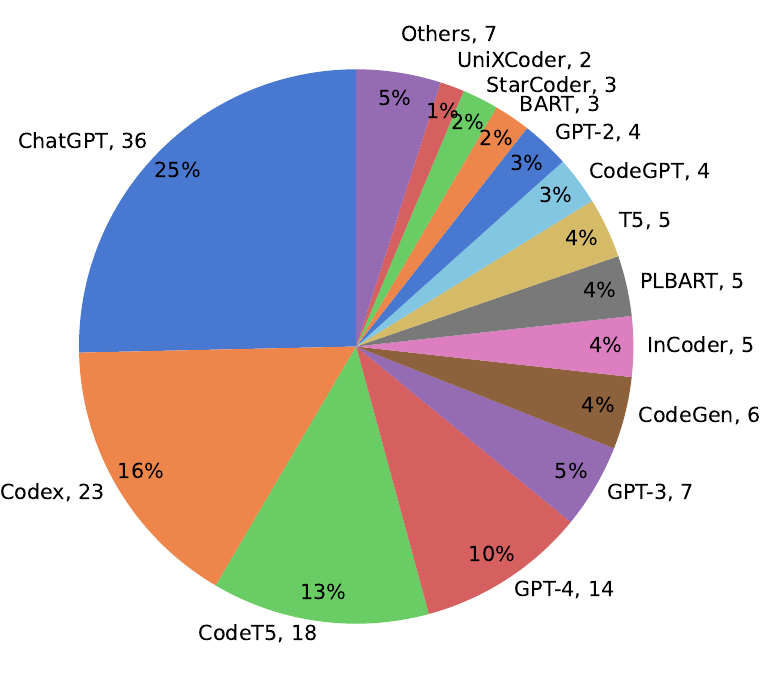}
\caption{LLMs used in the collected papers}
\label{fig:LLM_models}
\vspace{-0.1in}
\end{figure}

As shown in Figure \ref{fig:LLM_models}, the most commonly utilized LLM in software testing tasks is ChatGPT, which was released on Nov. 2022 by OpenAI. 
It is trained on a large corpus of natural language text data, and primarily designed for natural language processing and conversation. 
ChatGPT is the most widely recognized and popular LLM up until now, known for its exceptional performance across various tasks. Therefore, it comes as no surprise that it ranks in the top position in terms of our collected studies.


Codex, an LLM based on GPT-3, is the second most commonly used LLM in our collected studies.
It is trained on a massive code corpus containing examples from many programming languages such as JavaScript, Python, C/C++, and Java. 
Codex was released on Sep. 2021 by OpenAI and powers GitHub Copilot– an AI pair programmer that generates whole code snippets, given a natural language description as a prompt.
Since a large portion of our collected studies involve the source code (e.g., repair, unit test case generation), it is not surprising that researchers choose Codex as the LLM in assisting them in accomplishing the coding-related tasks. 

\begin{figure*}[t!]
\centering
\includegraphics[width=0.9\linewidth]{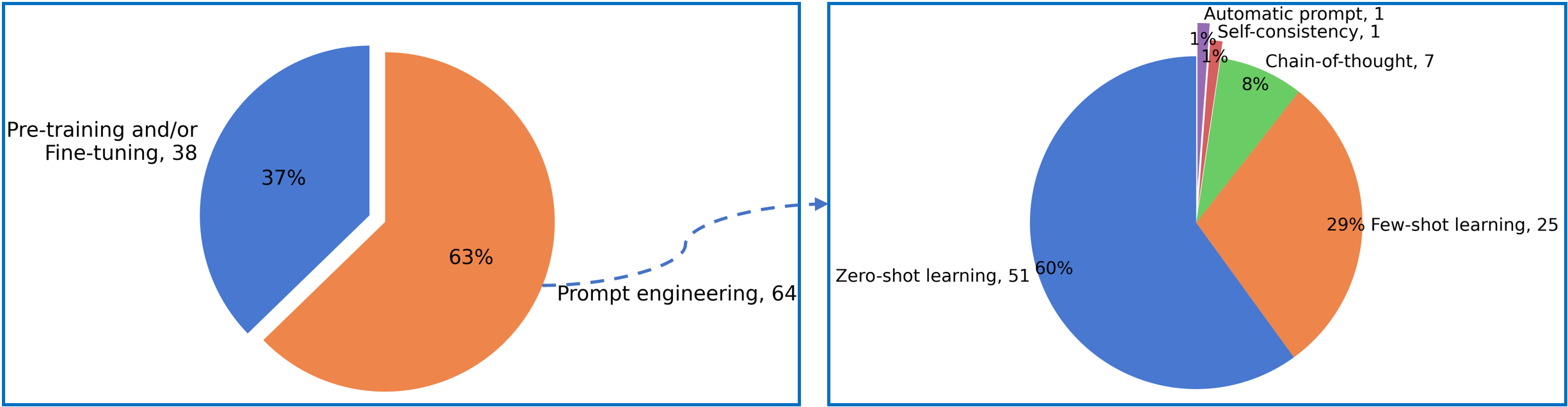}
\caption{Distribution about how LLM is used (Note that, a study can involve multiple types of prompt engineering) }

	

\label{fig:typeOfPrompt}
\vspace{-0.1in}
\end{figure*}

The third-ranked LLM is CodeT5, which is an open-sourced LLM developed by salesforce\footnote{https://blog.salesforceairesearch.com/codet5/}. 
Thanks to its open source, researchers can easily conduct the pre-training and fine-tuning with domain-specific data to achieve better performance. 
Similarly, CodeGen is also open-sourced and ranked relatively higher. 
Besides, for CodeT5 and CodeGen, there are more than half of the related studies involve the empirical evaluations (which employ multiple LLMs), e.g.,  program repair \cite{57howEffective,32impactLanguageModels}, unit test case generation \cite{66generatingUnitTests}.

There are already 14 studies that utilize GPT-4, ranking at the fourth place, which is launched on March 2023. 
Several studies directly utilize this state-of-the-art LLM of OpenAI, since it demonstrates excellent performance across a wide range of generation and reasoning tasks. 
For example, Xie et al. utilize GPT-4 to generate fuzzing inputs \cite{166Fuzz4allUniversalFuzzing}, while Vikram et al. employ it to generate property-based tests with the assistance of API documentation \cite{121CanLargeLanguage}.
In addition, some studies conduct experiments using both GPT-4 and ChatGPT or other LLMs to provide a more comprehensive evaluation of these models' performance. 
In their proposed LLM-empowered automatic penetration testing technique, Deng et al. find that GPT-4 surpasses ChatGPT and LaMDA from Google \cite{120PentestgptAnLlm}. 
Similarly, Zhang et al. find that GPT-4 shows its performance superiority over ChatGPT when generating the fuzz drivers with both the basic query strategies and enhanced query strategies \cite{140UnderstandingLargeLanguage}. 
Furthermore, GPT-4, as a multi-modal LLM, sets itself apart from the other mentioned LLMs by showcasing additional capabilities such as generating image narratives and answering questions based on images \cite{song2023howtobridge}.
Yet we have not come across any studies that explore the utilization of GPT-4's image-related features (e.g., UI screenshots, programming screencasts) in software testing tasks.





\subsection{Types of Prompt Engineering}
\label{subsec_LLM_prompt_type}

As shown in Figure \ref{fig:typeOfPrompt}, among our collected studies, 38 studies utilize the LLMs through pre-training or fine-tuning schema, while 64 studies employ the prompt engineering to communicate with LLMs to steer its behavior for desired outcomes without updating the model weights.
When using the early LLMs, their performances might not be as impressive, so researchers often use pre-training or fine-tuning techniques to adjust the models for specific domains and tasks in order to improve their performance. 
Then with the upgrading of LLM technology, especially with the introduction of GPT-3 and later LLMs, the knowledge contained within the models and their understanding/inference capability has increased significantly. 
Therefore, researchers will typically rely on prompt engineering to consider how to design appropriate prompts to stimulate the model's knowledge.

Among the 64 studies with prompt engineering, 51 studies involve zero-shot learning, and 25 studies involve few-shot learning (a study may involve multiple types). 
There are also studies involving the chain-of-though (7 studies), self-consistency (1 study), and automatic prompt (1 study). 

\textbf{Zero-shot learning} is to simply feed the task text to the model and ask for results.
Many of the collected studies employ the Codex, CodeT5, and CodeGen (as shown in Section \ref{subsec_LLM Models}), which is already trained on source code.
Hence, for the tasks dealing with source code like unit test case generation and program repair as demonstrated in previous sections, directly querying the LLM with prompts is the common practice.
There are generally two types of manners of zero-shot learning, i.e., with and without instructions. 
For example, Xie et al. \cite{63chatUniTest} would provide the LLMs with the instructions as ``please help me generate a JUnit test for a specific Java method ...'' to facilitate the unit test case generation. 
In contrast, Siddiq et al. \cite{66generatingUnitTests}  only provide the code header of the unit test case (e.g., ``class \$\{className\}\$\{suffix\}Test \{''), and the LLMs would carry out the unit test case generation automatically.
Generally speaking, prompts with clear instructions will yield more accurate results, while prompts without instructions are typically suitable for very specific situations.

\textbf{Few-shot learning} presents a set of high-quality demonstrations, each consisting of both input and desired output, on the target task. 
As the model first sees the examples, it can better understand human intention and criteria for what kinds of answers are wanted, which is especially important for tasks that are not so straightforward or intuitive to the LLM.
For example, when conducting the automatic test generation from general bug reports, Kang et al. \cite{9generalBugReproduction} provide examples of bug reports (questions) and the corresponding bug reproducing tests (answers) to the LLM, and their results show that two examples can achieve the highest performance than no examples or other number of examples. 
Another example of test assertion generation, Nashid et al. \cite{56retrievalPromptSelection} provide demonstrations of the focal method, the test method
containing an $<$AssertPlaceholder$>$, and the expected assertion, which enables the LLMs to better understand the task.

\textbf{Chain-of-thought (CoT) prompting} generates a sequence of short sentences to describe reasoning logics step by step (also known as reasoning chains or rationales) to the LLMs for generating the final answer. 
For example, for program repair from the natural language issue descriptions \cite{68codeEdits}, given the buggy code and issue report, the authors first ask the LLM to localize the bug, and then they ask it to explain why the localized lines are buggy, finally, they ask the LLM to fix the bug.
Another example is for generating unusual programs for fuzzing deep learning libraries, Deng et al. \cite{19fuzzDeepLearningLibraries} first generate a possible ``bug'' (bug description) before generating the actual ``bug-triggering'' code snippet that invokes the target API. 
The predicted bug description provides an additional hint to the LLM, indicating
that the generated code should try to cover specific potential buggy
behavior.

\textbf{Self-consistency} involves evaluating the coherence and consistency of the LLM's responses on the same input in different contexts.
There is one study with this prompt type, and it is about debugging. 
Kang et al. \cite{54explainableDebugging} employ a hypothesize-observe-conclude loop, which first generates a hypothesis about what the bug is and constructs an experiment to verify, using an LLM, then decide whether the hypothesis is correct based on the experiment result (with a debugger or code execution) using an LLM, after that, depending on the conclusion, it either starts with a new hypothesis or opts to terminate the debugging process and generate a fix.

\textbf{Automatic prompt} aims to automatically generate and select the appropriate instruction for the LLMs, instead of requiring the user to manually engineer a prompt.
Xia et al. \cite{166Fuzz4allUniversalFuzzing} introduce an auto-prompting step that automatically distils all user-provided inputs into a concise and effective prompt for fuzzing.
Specifically, they first generate a list of candidate prompts by incorporating the user inputs and auto prompting instruction while setting the LLM at high temperature, then a small-scale fuzzing experiment is conducted to evaluate each candidate prompt, and the best one is selected.

Note that there are fourteen studies that apply the iterative prompt design when using zero-shot or few-shot learning, in which the approach continuously refines the prompts with the running information of the testing task, e.g., the test failure information.  
For example, for program repair, Xia et al. \cite{2keepConversationGoing} interleave patch generation with test validation feedback to prompt future generation iteratively. 
In detail, they incorporate various information from a failing test including its name, the relevant code line(s) triggering the test failure, and the error message produced in the next round of prompting which can help the model understand the failure reason and provide guidance towards generating the correct fix.
Another example is for mobile GUI testing, Liu et al. \cite{60GUITesting} iteratively query the LLM about the operation (e.g., click a button, enter a text) to be conducted in the mobile app, and at each iteration, they would provide the LLM with current context information like which GUI pages and widgets have just explored. 

\begin{figure}[!ht]
\centering
\includegraphics[width=\linewidth]{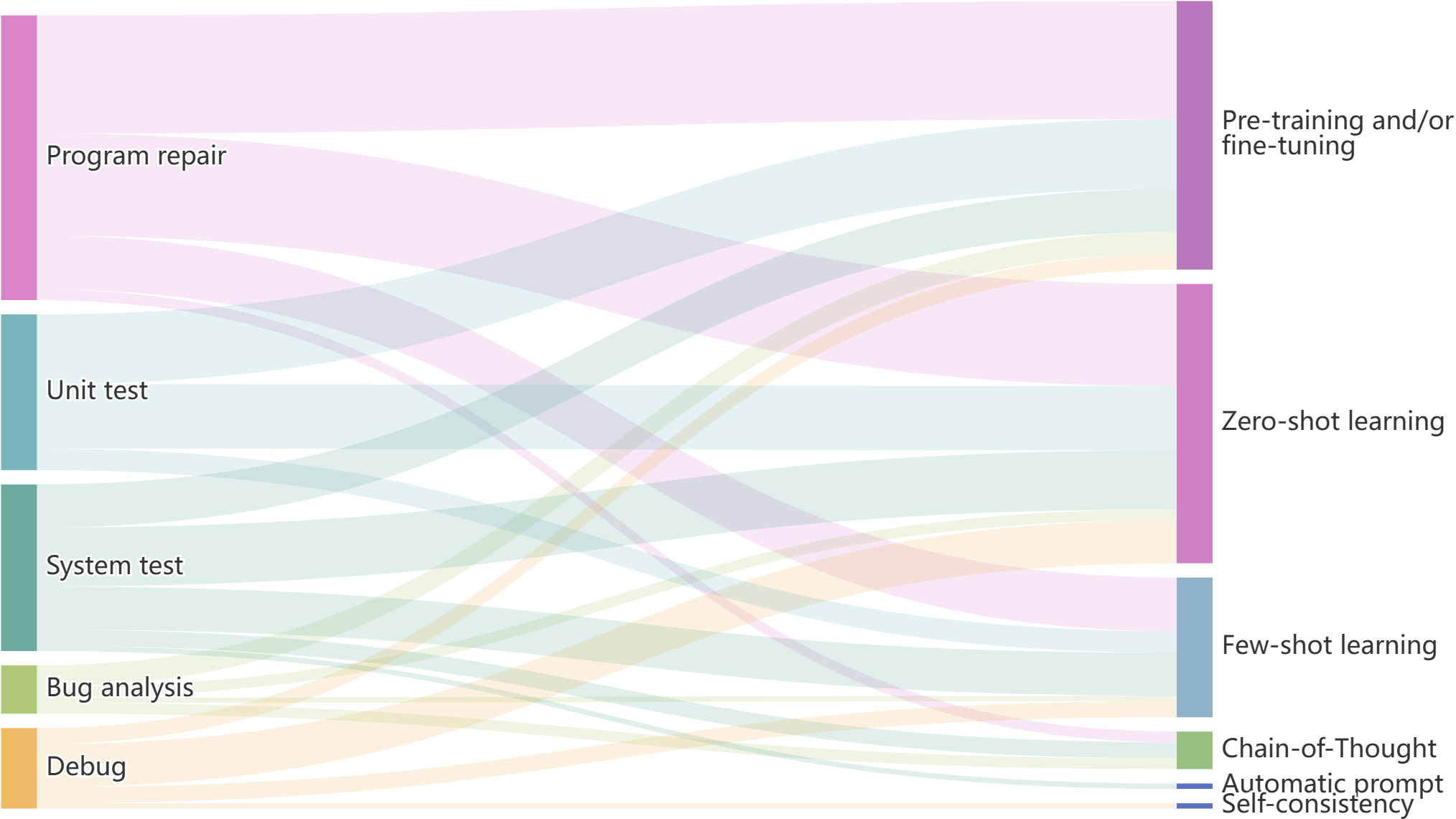}
\caption{Mapping between testing tasks and how LLMs are used}

\label{fig:taskPromptRelation}
\vspace{-0.1in}
\end{figure}

\textbf{Mapping between testing tasks and how LLMs are used.}
Figure \ref{fig:taskPromptRelation} demonstrates the mapping between the testing tasks (mentioned in Section \ref{sec_analysis_testing}) and how LLMs are used (as introduced in this subsection). 
The unit test case generation and program repair share similar patterns of communicating with the LLMs, since both tasks are closely related to the source code. 
Typically, researchers utilize pre-training and/or fine-tuning and zero-shot learning methods for these two tasks. 
Zero-shot learning is suitable because these tasks are relatively straightforward and can be easily understood by LLMs.
Moreover, since the training data for these two tasks can be automatically collected from source code repositories, pre-training and/or fine-tuning methods are widely employed for these two tasks, which can enhance LLMs' understanding of domain-specific knowledge.


In comparison, for system test input generation, zero-shot learning and few-shot learning methods are commonly used. 
This might be because this task often involves generating specific types of inputs, and demonstrations in few-shot learning can assist the LLMs in better understanding what should be generated. 
Besides, for this task, the utilization of pre-training and/or fine-tuning methods are not as widespread as in unit test case generation and program repair. 
This might be attributed to the fact that training data for system testing varies across different software and is relatively challenging to collect automatically.






\begin{figure}[t!]
\centering
\includegraphics[width=0.9\linewidth]{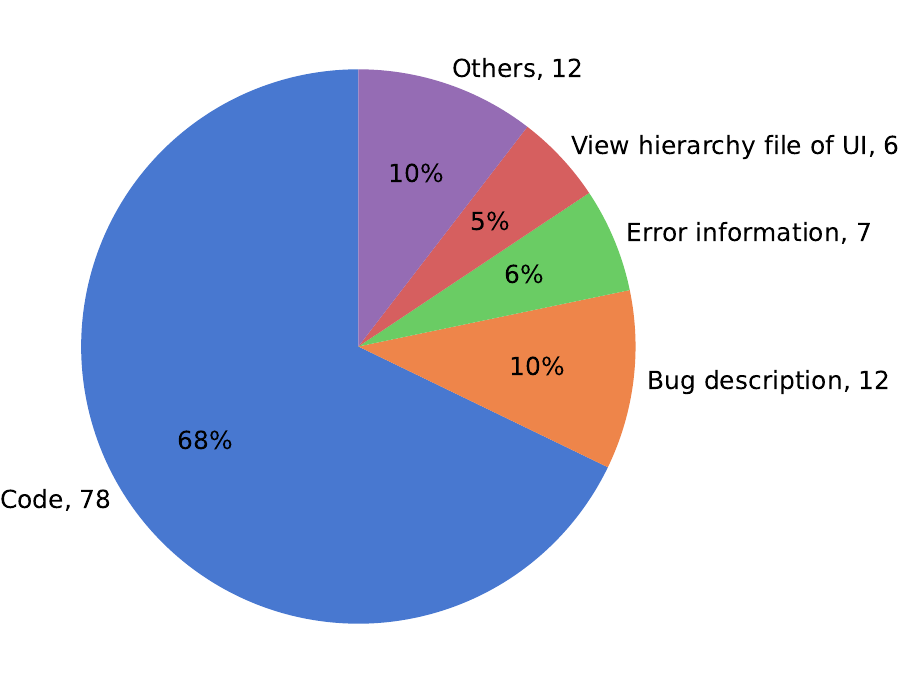}
\caption{Input of LLM}
\label{fig:inputOfLLM}
\vspace{-0.1in}
\end{figure}

\subsection{Input of LLM}
\label{subsec_LLM_input_output}

We also find that different testing tasks or software under test might involve diversified input when querying the LLM, as demonstrated in Figure \ref{fig:inputOfLLM}.

The most commonly utilized input is the \textbf{source code} since a large portion of collected studies relate to program repair or unit test case generation whose input are source code. 
For unit test case generation, typical code-related information would be (i) the complete focal method, including the signature and
body; (ii) the name of the focal class (i.e., the class that the
focal method belongs to); (iii) the field in the focal class; and
(iv) the signatures of all methods defined in the focal class \cite{28unitTest,64noUnitTest}.
For program repair, there can be different setups and involve different inputs, including (i) inputting a buggy function with the goal of outputting the patched function, (ii) inputting the buggy location with the goal of generating the correct replacement code (can be a single line change) given the prefix and suffix of the buggy function \cite{33practicalProgramRepair}.
Besides, there can be variations for the buggy location input, i.e., (i) does not contain the buggy lines (but the bug
location is still known), (ii) give the buggy lines as lines of comments.

There are also 12 studies taking the \textbf{bug description} as input for the LLM.
For example, Kang et al. \cite{9generalBugReproduction} take the bug description as input when querying LLM and let the LLM generate the bug-reproducing test cases. Fakhoury et al. \cite{68codeEdits} input the natural language
descriptions of bugs to the LLM, and generate the correct code fixes. 

There are 7 studies that would provide the \textbf{intermediate error information}, e.g., test failure information, to the LLM, and would conduct the iterative prompt (as described in Section \ref{subsec_LLM_prompt_type}) to enrich the context provided to the LLM. 
These studies are related to the unit test case generation and program repair, since in these scenarios, the running information can be acquired easily.

When testing mobile apps, since the utilized LLM could not understand the image of the GUI page, the \textbf{view hierarchy file} which represents the details of the GUI page usually acts as the input to LLMs.  
Nevertheless, with the emergence of GPT-4 which is a multimodal model and accepts both image and text inputs for model input, the GUI screenshots might be directly utilized for LLM's input.




\subsection{Incorporating Other Techniques with LLM}
\label{subsec_LLM+X}

\begin{figure*}[t!]
\centering
\includegraphics[width=0.9\linewidth]{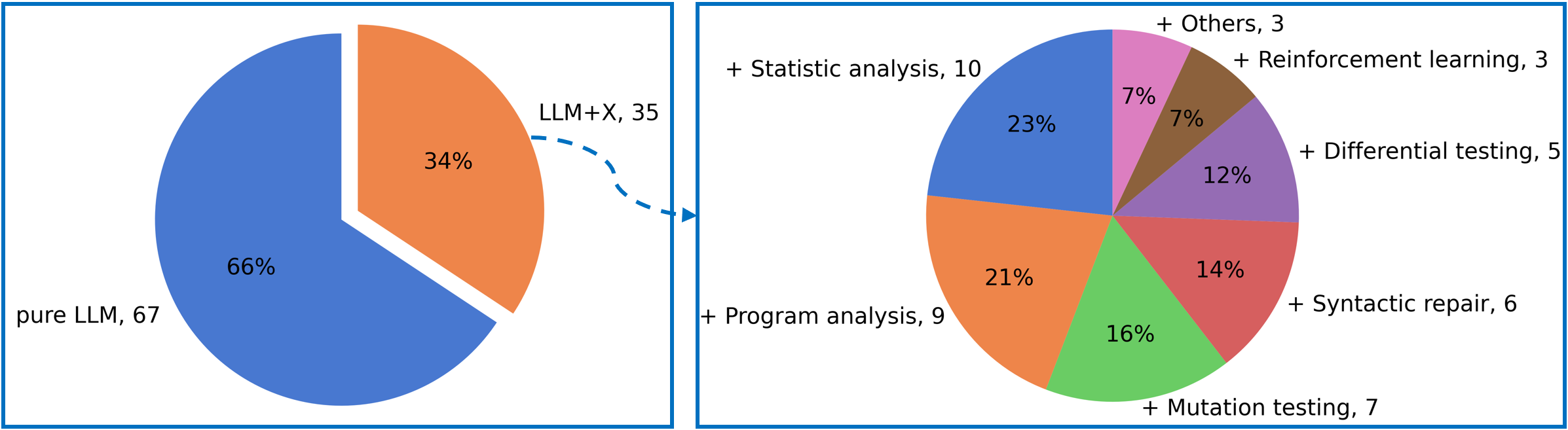}
\caption{Distribution about other techniques incorporated with LLMs (Note that, a study can involve multiple types) }

\label{fig:LLMandX}
\vspace{-0.1in}
\end{figure*}

There are divided opinions on whether LLM has reached an all-powerful status that requires no other techniques. 
As shown in Figure \ref{fig:LLMandX}, among our collected studies, 67 of them utilize LLMs to address the entire testing task, while 35 studies incorporate additional techniques. 
These techniques include mutation testing, differential testing, syntactic checking, program analysis, statistical analysis, etc. 
.

The reason why researchers still choose to combine LLMs with other techniques might be because, despite exhibiting enormous potential in various tasks, LLMs still possess limitations such as comprehending code semantics and handling complex program structures. 
Therefore, combining LLMs with other techniques optimizes their strengths and weaknesses to achieve better outcomes in specific scenarios. 
In addition, it is important to note that while LLMs are capable of generating correct code, they may not necessarily produce sufficient test cases to check for edge cases or rare scenarios. 
This is where mutation and other testing techniques come into play, as they allow for the generation of more diverse and complex code that can better simulate real-world scenarios. 
Taken in this sense, a testing approach can incorporate a combination of different techniques, including both LLMs and other testing strategies, to ensure comprehensive coverage and effectiveness.



\textbf{LLM + statistical analysis.}
As LLMs can often generate a multitude of outputs, manually sifting through and identifying the correct output can be overwhelmingly laborious. 
As such, researchers have turned to statistical analysis techniques like ranking and clustering \cite{9generalBugReproduction,17interactiveCodeGeneration,24semanticsTestCompletion,33practicalProgramRepair,71repairProgramDependence} to efficiently filter through LLM's outputs and ultimately obtain more accurate results.

\textbf{LLM + program analysis.} 
When utilizing LLMs to accomplish tasks such as generating unit test cases and repairing software code, it is important to consider that software code inherently possesses structural information, which may not be fully understood by LLMs.
Hence, researchers often utilize program analysis techniques, including code abstract syntax trees (ASTs) \cite{5explainingSoftwareBugs}, to represent the structure of code more effectively and increase the LLM's ability to comprehend the code accurately. 
Researchers also perform the structure-based subsetting of code lines to narrow the focus for LLM \cite{8bugsPythonAssignments}, or extract additional code context from other code files \cite{64noUnitTest}, to enable the models to focus on the most task-relevant information in the codebase and lead to more accurate predictions. 

\textbf{LLM + mutation testing.} It is mainly targeting at generating more diversified test inputs. 
For example, Deng et al. \cite{13fuzzDeepLearningLibraries} first use LLM to generate the seed programs (e.g., code snippets using a target DL API) for fuzzing deep learning libraries. 
To enrich the pool of these test programs, they replace parts of the seed program with masked tokens using mutation operators (e.g., replaces the API call arguments with the span token) to produce masked inputs, and again utilize the LLMs to perform code infilling to generate new code that replaces the masked tokens.

\textbf{LLM + syntactic checking.} 
Although LLMs have shown remarkable performance in various natural language processing tasks, the generated code from these models can sometimes be syntactically incorrect, leading to potential errors and reduced usability. 
Therefore, researchers have proposed to leverage syntax checking to identify and correct errors in the generated code.
For example, in their work for unit test case generation, Alagarsamy et al. \cite{23a3testTestGeneration} additionally introduce a verification method to check and repair the naming consistency (i.e., revising the test method name to be consistent with the focal method name) and the test signatures (i.e., adding missing keywords like public, void, or @test annotations).
Xie et al. \cite{63chatUniTest} also validates the generated unit test case and employs rule-based repair to fix syntactic and simple compile errors.

\textbf{LLM + differential testing.} Differential testing is well-suited to find semantic or logic bugs that do not exhibit explicit erroneous behaviors like crashes or assertion failures. 
In this category of our collected studies, the LLM is mainly responsible for generating valid and diversified inputs, while the differential testing helps to determine whether there is a triggered bug based on the software's output. 
For example, Ye et al. \cite{14conformanceTesting} first uses LLM to produce random JavaScript programs, and leverages the language specification document to generate test data, then conduct the differential testing on JavaScript engines such as JavaScriptCore, ChakraCore, SpiderMonkey, QuickJS, etc. 
There are also studies utilizing the LLMs to generate test inputs and then conduct differential testing for fuzzing DL libraries 
\cite{13fuzzDeepLearningLibraries,19fuzzDeepLearningLibraries} and SAT solvers \cite{204SmtSolverValidation}. 
Li et al. \cite{67failureTestCases} employs the LLM in finding the failure-inducing test cases. 
In detail, given a program under test, they first request the LLM to infer the intention of the program, then request the LLM to generate programs that have the same intention, which are alternative implementations of the program, and are likely free of the program's bug.
Then they perform the differential testing with the program under test and the generated programs to find the failure-inducing test cases.



\section{Challenges and Opportunities}
\label{sec_challenge_opportunity}

Based on the above analysis from the viewpoints of software testing and LLM, we summarize the challenges and opportunities when conducting software testing with LLM. 


\subsection{Challenges}
\label{subsec_challenge}

As indicated by this survey, software testing with LLMs has undergone significant growth in the past two years. 
However, it is still in its early stages of development, and numerous challenges and open questions need to be addressed.


\subsubsection{Challenges for Achieving High Coverage}
Exploring the diverse behaviors of the software under test to achieve high coverage is always a significant concern in software testing. 
In this context, test generation differs from code generation, as code generation primarily focuses on producing a single, correct code snippet, whereas software testing requires generating diverse test inputs to ensure better coverage of the software. 
Although setting a high temperature can facilitate the LLMs in generating different outputs, it remains challenging for LLMs to directly achieve the required diversity. 
For example, for unit test case generation, in SF110 dataset, the line coverage is merely 2\% and the branch coverage is merely 1\% \cite{66generatingUnitTests}.
For system test input generation, in terms of fuzzing DL libraries, the API coverage for TensorFlow is reported to be 66\% (2215/3316) \cite{13fuzzDeepLearningLibraries}.


From our collected studies, we observe that the researchers often utilize mutation testing together with the LLMs to generate more diversified outputs. 
For example, when fuzzing a DL library, instead of directly generating the code snippet with LLM, Deng et al. \cite{13fuzzDeepLearningLibraries} replace parts of the selected seed (code generated by LLM) with masked tokens using different mutation operators to produce masked inputs. They then leverage the LLM to perform code infilling to generate new code that replaces the masked tokens, which can significantly increase the diversity of the generated tests. 
Liu et al. \cite{107TestingTheLimits} leverage LLM to produce the test generators (each of which can yield a batch of unusual text inputs under the same mutation rule) together with the mutation rules for text-oriented fuzzing, which reduces the human effort required for designing mutation rules.

A potential research direction could involve utilizing testing-specific data to train or fine-tune a specialized LLM that is specifically designed to understand the nature of testing. 
By doing so, the LLM can inherently acknowledge the requirements of testing and autonomously generate diverse outputs. 


\subsubsection{Challenges in Test Oracle Problem}
The oracle problem has been a longstanding challenge in various testing applications, e.g., testing machine learning systems \cite{zhang2022machineLearningTesting} and testing deep learning libraries \cite{13fuzzDeepLearningLibraries}.
To alleviate the oracle problem to the overall testing activities, a common practice in our collected studies is to transform it into a more easily derived form, often by utilizing differential testing \cite{204SmtSolverValidation} or focusing on only identifying crash bugs \cite{60GUITesting}. 

There are successful applications of differential testing with LLMs, as shown in Figure \ref{fig:LLMandX}. 
For instance,  
when testing the SMT solvers, Sun et al. adopt differential
testing which involves comparing the results of multiple SMT
solvers (i.e., Z3, cvc5, and Bitwuzla) on the same generated test formulas by LLM \cite{204SmtSolverValidation}.
However, this approach is limited to systems where counterpart software or running environment can easily be found, potentially restricting its applicability. 
Moreover, to mitigate the oracle problem, other studies only focus on the crash bugs which are easily observed automatically.  
This is particularly the case for mobile applications testing, in which the LLMs guide the testing in exploring more diversified pages, conducting more complex operational actions, and covering more meaningful operational sequences \cite{60GUITesting}.
However, this significantly restricts the potential of utilizing the LLMs for uncovering various types of software bugs.

Exploring the use of LLMs to derive other types of test oracles represents an interesting and valuable research direction. 
Specifically, metamorphic testing is also widely used in software testing practices to help mitigate the oracle problem, yet in most cases, defining metamorphic relations relies on human ingenuity. 
Luu et al. \cite{106CanChatgptAdvance} have examined the effectiveness of LLM in generating metamorphic relations, yet they only experiment with straightforward prompts by directly querying ChatGPT.
Further exploration, potentially incorporating human-computer interaction or domain knowledge, is highly encouraged. 
Another promising avenue is exploring the capability of LLMs to automatically generate test cases based on metamorphic relations, covering a wide range of inputs. 

The advancement of multi-model LLMs like GPT-4 may open up possibilities for exploring their ability to detect bugs in software user interfaces and assist in deriving test oracles.
By leveraging the image understanding and reasoning capabilities of these models, one can investigate their potential to automatically identify inconsistencies, errors, or usability issues in user interfaces. 

\subsubsection{Challenges for Rigorous Evaluations}
The lack of benchmark datasets and the potential data leakage issues associated with LLM-based techniques present challenges in conducting rigorous evaluations and comprehensive comparisons of proposed methods. 

For program repair, there are only two well-known and commonly-used benchmarks, i.e., Defect4J and QuixBugs, as demonstrated in Table \ref{tab:repair}.
Furthermore, these datasets are not specially designed for testing the LLMs. 
For example, as reported by Xia et al. \cite{33practicalProgramRepair}, 39 out of 40 Python bugs in the QuixBugs dataset can be fixed by Codex, yet in real-world practice, the successful fix rate can be nowhere near as high.
For unit test case generation, there are no widely recognized benchmarks, and different studies would utilize different datasets for performance evaluation, as demonstrated in Table \ref{tab:unit_test}. 
This indicates the need to build more specialized and diversified benchmarks.

Furthermore, the LLMs may have seen the widely-used benchmarks in their pre-training data, i.e., data leakage issues. 
Jiang et al. \cite{32impactLanguageModels} check the CodeSearchNet and BigQuery, which are the data sources of common LLMs, and the results show that four repositories used by the Defect4J benchmark are also in CodeSearchNet, and the whole Defects4J repository is included by BigQuery. 
Therefore, it is very likely that existing program repair benchmarks are seen by the LLMs during pre-training. 
This data leakage issue has also been investigated in machine learning-related studies. For example, Tu et al. \cite{tu2018becareful} focus on the data leakage in issue tracking data, and results show that information leaked from the ``future'' makes prediction models misleadingly optimistic. 
This reminds us that the performance of LLMs on software testing tasks may not be as good as reported in previous studies. 
It also suggests that we need more specialized datasets that are not seen by LLMs to serve as benchmarks.
One way is to collect it from specialized sources, e.g., user-generated content from niche online communities.

\subsubsection{Challenges in Real-world Application of LLMs in Software Testing}
As we mentioned in Section \ref{subsec_LLM_prompt_type}, in the early days of using LLMs, pre-training and fine-tuning are commonly used practice, considering the model parameters are relatively few resulting in weaker model capabilities (e.g., T5).
As time progressed, the number of model parameters increased significantly, leading to the emergence of models with greater capabilities (e.g., ChatGPT). 
And in recent studies, prompt engineering has become a common approach. 
However, due to concerns regarding data privacy, when considering real-world practice, most software organizations tend to avoid using commercial LLMs and would prefer to adopt open-source ones with training or fine-tuning using organization-specific data. 
Furthermore, some companies also consider the current limitations in terms of computational power or pay close attention to energy consumption, they tend to fine-tune medium-sized models.
It is quite challenging for these models to achieve similar performance to what our collected papers have reported. 
For instance, in the widely-used QuixBugs dataset, it has been reported that 39 out of 40 Python bugs and 34 out of 40 Java bugs can be automatically fixed \cite{33practicalProgramRepair}. 
However, when it comes to DL programs collected from Stack Overflow, which represent real-world coding practice, only 16 out of 72 Python bugs can be automatically fixed \cite{52promptDesign}.

Recent research has highlighted the importance of high-quality training data in improving the performance of models for code-related tasks \cite{sun2022ontheimportance}, yet manually building high-quality organization-specific datasets for training or fine-tuning is time-consuming and labor-intensive.
To address this, one is encouraged to utilize the automated techniques of mining software repositories to build the datasets, for example, techniques like key information extraction techniques from Stack Overflow \cite{shi2021ispy} offer potential solutions for automatically gathering relevant data.  

In addition, exploring the methodology for better fine-tuning the LLMs with software-specific data is worth considering because software-specific data differs from natural language data as it contains more structural information, such as data flow and control flow. 
Previous research on code representations has shown the benefits of incorporating data flow, which captures the semantic-level structure of code and represents the relationship between variables in terms of ``whether-value-comes-from'' \cite{guo2021graphcodebert}. 
These insights can provide valuable guidance for effectively fine-tuning LLMs with software-specific data.

\subsection{Opportunities}
\label{subsec_opportunities}

There are also many research opportunities in software testing with LLMs, which can greatly benefit developers, users, and the research community. 
While not necessarily challenges, these opportunities contribute to advancements in software testing, benefiting practitioners and the wider research community.

\subsubsection{Exploring LLMs in the Early Stage of Testing}
As shown in Figure \ref{fig:testingTasks}, LLMs have not been used in the early stage of testing, e.g., test requirements, and test planning.
There might be two main reasons behind that. 
The first is the subjectivity in early-stage testing tasks.
Many tasks in the early stages of testing, such as requirements gathering, test plan creation, and design reviews, may involve subjective assessments that require significant input from human experts. This could make it less suitable for LLMs that rely heavily on data-driven approaches.
The second might be the lack of open-sourced data in the early stages. Unlike in later stages of testing, there may be limited data available online during early-stage activities. This could mean that LLMs may not have seen much of this type of data, and therefore may not perform well on these tasks.

Adopting a human-computer interaction schema for tackling early-stage testing tasks would harness the domain-specific knowledge of human developers and leverage the general knowledge embedded in LLMs. 
Additionally, it is highly encouraged for software development companies to record and provide access to early-stage testing data, allowing for improved training and performance of LLMs in these critical testing activities.

\subsubsection{Exploring LLMs in Other Testing Phases}
\begin{figure}[!ht]
\centering
\includegraphics[width=\linewidth]{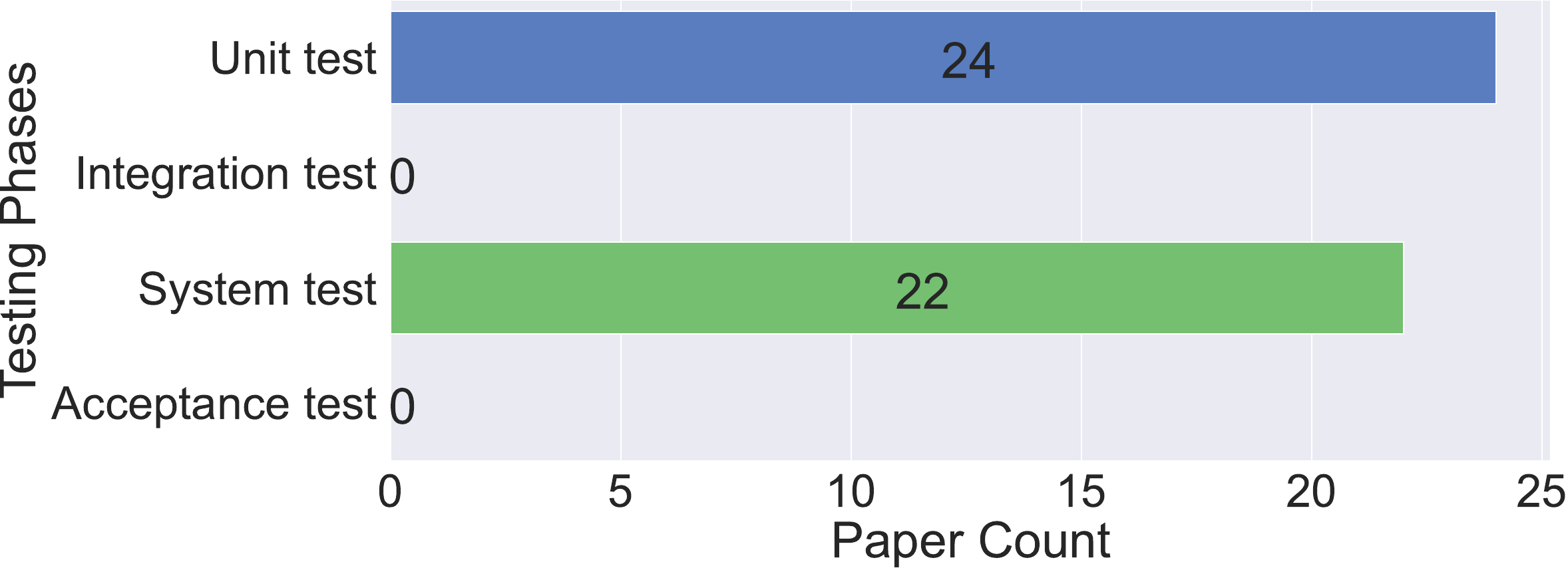}
\caption{Distribution of testing phases (note that we omit the studies which do not explicitly specify the testing phases, e.g., program repair)}

\label{fig:testingPhase}
\vspace{-0.1in}
\end{figure}

We have analyzed the distribution of testing phases for the collected studies.
As shown in Fig \ref{fig:testingPhase}, we can observe that LLMs are most commonly used in unit testing, followed by system testing. However, there is still no research on the use of LLMs in integration testing and acceptance testing.




For integration testing, it involves testing the interfaces between different software modules.
In some software organizations, integration testing might be merged with unit testing, which can be a possible reason why LLM is rarely utilized in integration testing.
Another reason might be that the size and complexity of the input data in this circumstance may exceed the capacity of the LLM to process and analyze (e.g., the source code of all involved software modules), which can lead to errors or unreliable results.  
To tackle this, a potential reference can be found in Section \ref{subsec_testingtask_unit}, where Xie et al. \cite{63chatUniTest} design a method to organize the necessary information into the pre-defined maximum prompt token limit of the LLM.
Furthermore, integration testing requires diversified data to be generated to sufficiently test the interface among multiple modules. 
As mentioned in Section \ref{subsec_testingtask_input}, previous work has demonstrated the LLM's capability in generating diversified test input for system testing, in conjunction with mutation testing techniques \cite{13fuzzDeepLearningLibraries,14conformanceTesting}.
And these can provide insights about generating the diversified interface data for integration testing. 


Acceptance testing is usually conducted by business analysts or end-users to validate the system's functionality and usability, which requires more non-technical language and domain-specific knowledge, thus making it challenging to apply LLM effectively.
Since acceptance testing involves humans, it is well-suited for the use of human-in-the-loop schema with LLMs. 
This has been studied in traditional machine learning \cite{yu2015lsun}, but has not yet been explored with LLMs. Specifically, the LLMs can be responsible for automatically generating test cases, evaluating test coverage, etc, while human testers are responsible for checking the program's behavior and verifying test oracle.



\subsubsection{Exploring LLMs for More Types of Software}
We analyze what types of software have been explored in the collected studies, as shown in Figure \ref{fig:softwareUnderTest}.
Note that, since a large portion of studies are focused on unit testing or program repair, they are conducted on publicly available datasets and do not involve specific software types.

From the analysis in Section \ref{subsec_testingtask_input}, the LLM can generate not only the source code for testing DL libraries but also the textual input for testing mobile apps, even the models for testing CPS. 
Overall, the LLM provides a flexible and powerful framework for generating test inputs for a wide range of applications. 
Its versatility would make it useful for testing the software in other domains.

From one point of view, some proposed techniques can be applied to other types of software. 
For example, in the paper proposed for testing deep learning libraries \cite{19fuzzDeepLearningLibraries}, since it proposes techniques for generating diversified, complicated, and human-like DL programs, the authors state that the approach can be easily extended to test software systems from other application domains, e.g., interpreters, database systems, and other popular libraries. 
More than that, there are already studies that focus on universal fuzzing techniques \cite{166Fuzz4allUniversalFuzzing,141AugmentingGreyboxFuzzing} which are designed to be adaptable and applicable to different types of test inputs and software.


From another point of view, other types of software can also benefit from the capabilities of LLMs to design the testing techniques that are better suited to their specific domain and characteristics. 
For instance, the metaverse, with its immersive virtual environments and complex interactions, presents unique challenges for software testing.
LLMs can be leveraged to generate diverse and realistic inputs that mimic user behavior and interactions within the metaverse, which are never explored.



\subsubsection{Exploring LLMs for Non-functional Testing}
In our collected studies, LLMs are primarily used for functional testing, and no practice in performance testing, usability testing or others.
One possible reason for the prevalence of LLM-based solutions in functional testing is that they can convert functional testing problems into code generation or natural language generation problems \cite{13fuzzDeepLearningLibraries,60GUITesting}, which LLMs are particularly adept at solving. 






On the other hand, performance testing and usability testing may require more specialized models that are designed to detect and analyze specific types of data, handle complex statistical analyses, or determine the buggy criteria. 
Moreover, there have been dozens of performance testing tools (e.g., LoadRunner \cite{Loadrunner}) that can generate a workload that simulates real-world usage scenarios and achieve relatively satisfactory performance. 

The potential opportunities might let the LLM integrate the performance testing tools and acts like the LangChain \cite{LangChain}, to better simulate different types of workloads based on real user behavior.
Furthermore, the LLMs can identify the parameter combinations and values that have the highest potential to trigger performance problems. 
It is essentially a way to rank and prioritize different parameter settings based on their impact on performance and improve the efficiency of performance testing.

\subsubsection{Exploring Advanced Prompt Engineering}
\begin{figure}[h!]
\centering
\includegraphics[width=\linewidth]{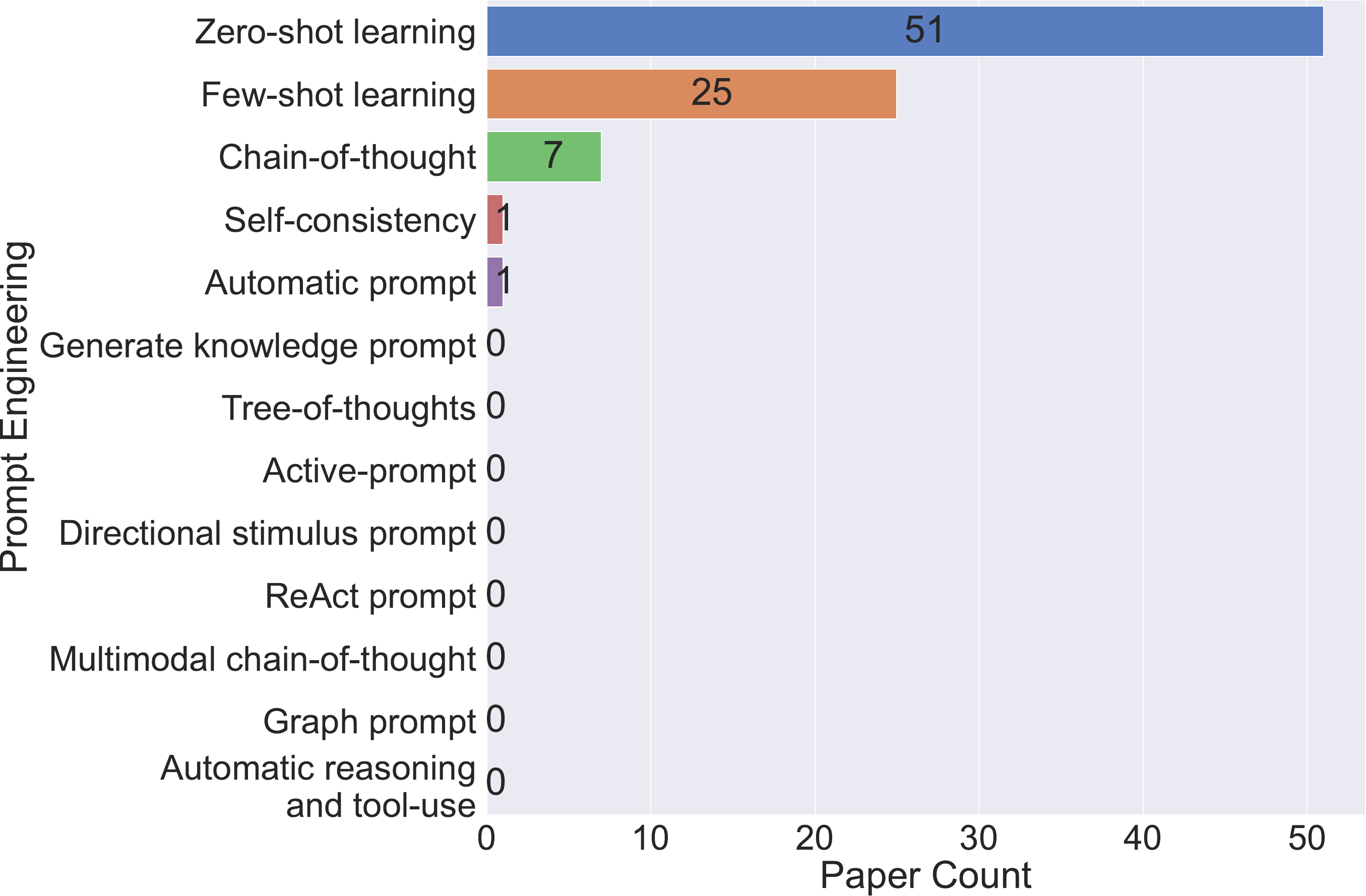}
\caption{List of advanced prompt engineering practices and those utilized in the collected papers}
\label{fig:promptEngineering}
\vspace{-0.1in}
\end{figure}

There are a total of 11 commonly used prompt engineering techniques as listed in a popular prompt engineering guide \cite{Promptengineeringguide}, as shown in Figure \ref{fig:promptEngineering}. 
Currently, in our collected studies, only the first five techniques are being utilized. 
The more advanced techniques have not been employed yet, and can be explored in the future for prompt design.




For instance, multimodal chain of thought prompting involves using diverse sensory and cognitive cues to stimulate thinking and creativity in LLMs \cite{Zhang2023multimodalCOT}. 
By providing images (e.g., GUI screenshots) or audio recordings related to the software under test can help the LLM better understand the software's context and potential issues. 
Besides, try to prompt the LLM to imagine itself in different roles, such as a developer, user, or quality assurance specialist. This perspective-shifting exercise enables the LLM to approach software testing from multiple viewpoints and uncover different aspects that might require attention or investigation.

Graph prompting \cite{liu2023GraphPrompt} involves the representation of information using graphs or visual structures to facilitate understanding and problem-solving. 
Graph prompting can be a natural match with software engineering, consider it involves various dependencies, control flow, data flow, state transitions, or other relevant graph structure. 
Graph prompting can be beneficial in analyzing this structural information, and enabling the LLMs to comprehend the software under test effectively. 
For instance, testers can use graph prompts to visualize test coverage, identify untested areas or paths, and ensure adequate test execution.


\subsubsection{Incorporating LLMs with Traditional Techniques}
There is currently no clear consensus on the extent to which LLMs can solve software testing problems. 
From the analysis in Section \ref{subsec_LLM+X}, we have seen some promising results from studies that have combined LLMs with traditional software testing techniques. 
This implies the LLMs are not the sole silver bullet for software testing. 
Considering the availability of many mature software testing techniques and tools, and the limited capabilities of LLMs, it is necessary to explore other better ways to combine LLMs with traditional testing or program analysis techniques and tools for better software testing.

Based on the collected studies, the LLMs have been successfully utilized together with various techniques such as differential testing (e.g., \cite{204SmtSolverValidation}), mutation testing (e.g., \cite{13fuzzDeepLearningLibraries}), program analysis (e.g., \cite{DomainKnowledge}, as shown in Figure \ref{fig:LLMandX}. 
From one perspective, future studies can explore improved integration of these traditional techniques with LLMs. 
Take mutation testing as an example, current practices mainly rely on the human-designed mutation rules to mutate the candidate tests, and let the LLMs re-generate new tests \cite{13fuzzDeepLearningLibraries,117EffectiveTestGeneration,166Fuzz4allUniversalFuzzing}, while Liu et al. directly utilize the LLMs for producing the mutation rules alongside the mutated tests \cite{107TestingTheLimits}.
Further explorations in this direction are of great interest.

From another point of view, more traditional techniques can be incorporated in LLMs for software testing. 
For instance, besides the aforementioned traditional techniques, the LLMs have been combined with formal verification for self-healing software detection in the field of software security \cite{charalambous2023new}.
More attempts are encouraged. 
Moreover, considering the existence of numerous mature software testing tools, one can explore the integration of LLMs with these tools, allowing them to act as a ``LangChain'' to better explore the potential of these tools. 



\section{Related Work}
\label{sec_related}


The systematic literature review is a crucial manner for gaining insights into the current trends and future directions within a particular field. 
It enables us to understand and stay updated on the developments in that domain.

Wang et al. surveyed the machine learning and deep learning techniques for software engineering \cite{wang2023MLDLforSE}.
Yang et al. and Watson et al. respectively carried out surveys about the use of deep learning in software engineering domain \cite{yang2022surveyDLforSE,Watson2022surveyDLforSE}.
Bajammal et al. surveyed the utilization of computer vision techniques to improve software engineering tasks \cite{Bajammal2022surveyCVforSE}.
Zhang et al. provided a survey of techniques for testing machine learning systems \cite{zhang2022machineLearningTesting}

With the advancements of artificial intelligence and LLMs, researchers also conduct systematic literature reviews about LLMs, and their applications in various fields (e.g., software engineering). 
Zhao et al. \cite{zhao2023surveyLLM} reviewed recent advances in LLMs by providing an overview of their background, key findings, and mainstream techniques. 
They focused on four major aspects of LLMs, namely pre-training, adaptation tuning, utilization, and capacity evaluation. 
Additionally, they summarized the available resources for developing LLMs and discuss the remaining issues for future directions.
Hou et al. conducted a systematic literature review on using LLMs for software engineering, with a particular
focus on understanding how LLMs can be exploited to optimize processes and outcomes \cite{hou2023surveyLLMforSE}.
Fan et al. conducted a survey of LLMs for software engineering, and set out open research challenges for the application of LLMs to technical problems faced by software engineers \cite{fan2023surveyLLMforSE}.
Zan et al. conducted a survey of existing LLMs for NL2Code task (i.e., generating code from a natural language description), and reviewed  benchmarks
and metrics \cite{zan2023surveyNL2code}.

While these studies either targeted the broader software engineering domain (with a limited focus on software testing tasks) or focused on other software development tasks (excluding software testing), this paper specifically focuses on the use of LLMs for software testing. 
It surveys related studies, summarizes key challenges and potential opportunities, and serves as a roadmap for future research in this area.

\section{Conclusion}
\label{sec_conclusion}


This paper provides a comprehensive review of the use of LLMs in software testing. We have analyzed relevant studies that have utilized LLMs in software testing from both the software testing and LLMs perspectives. 
This paper also highlights the challenges and potential opportunities in this direction. 
Results of this review demonstrate that LLMs have been successfully applied in a wide range of testing tasks, including unit test case generation, test oracle generation, system test input generation, program debugging, and program repair.
However, challenges still exist in achieving high testing coverage, addressing the test oracle problem, conducting rigorous evaluations, and applying LLMs in real-world scenarios.
Additionally, it is observed that LLMs are commonly used in only a subset of the entire testing lifecycle, for example, they are primarily utilized in the middle and later stages of testing, only serving the unit and system testing phases, and only for functional testing. 
This highlights the research opportunities for exploring the uncovered areas.
Regarding how the LLMs are utilized, we find that various pre-training/fine-tuning and prompt engineering methods have been developed to enhance the capabilities of LLMs in addressing testing tasks. 
However, more advanced techniques in prompt design have yet to be explored and can be an avenue for future research.

It can serve as a roadmap for future research in this area, identifying gaps in our current understanding of the use of LLMs in software testing and highlighting potential avenues for exploration. 
We believe that the insights provided in this paper will be valuable to both researchers and practitioners in the field of software engineering, assisting them in leveraging LLMs to improve software testing practices and ultimately enhance the quality and reliability of software systems.

\bibliographystyle{IEEEtran}
\normalem
\bibliography{paper, reference}


\newpage
\begin{sidewaystable*}[!t]
\footnotesize
\caption{All details of the collected papers}
\label{tab:collectedPaperDetailed}
\begin{tabular}{p{0.3cm}|p{5.5cm}|p{0.5cm}|p{2.5cm}|p{2.5cm}|p{3cm}|p{2cm}|p{2.5cm}|p{1.5cm}|p{0.5cm}}
\hline
\textbf{ID} & \textbf{Paper title} & \textbf{Year} & \textbf{Topic} & \textbf{Involved  LLM} & \textbf{How LLM is used} & \textbf{Input to LLM} & \textbf{How LLM integrated} & \textbf{Venue} & \textbf{Ref} \\ \hline
1 & Unit Test Case Generation with   Transformers and Focal Context & 2021 & Unit test case generation & BART & Pre-training and/or Fine-tuning & \textbf{Code} & \textbf{Pure LLM} & \textbf{Arxiv} & \textbf{\cite{28unitTest}} \\
2 & Codet: Code Generation with Generated Tests & 2022 & Unit test case generation & Codex & Zero-shot learning & Code & Pure LLM & ICLR 2023 & \cite{18codetGeneratedTests} \\
3 & Interactive Code Generation via Test-Driven User-Intent   Formalization & 2022 & Unit test case generation & Codex & Zero-shot learning & Code & Mutation testing; Statistic analysis & Arxiv & \cite{17interactiveCodeGeneration} \\
4 & A3Test: Assertion-Augmented Automated Test Case Generation & 2023 & Unit test case generation & PLBART & Pre-training and/or Fine-tuning & Code & Syntactic repair & Arxiv & \cite{23a3testTestGeneration} \\
5 & An Empirical Evaluation of Using Large Language Models for   Automated Unit Test Generation & 2023 & Unit test case generation & ChatGPT & Zero-shot learning & Code; Others & Syntactic repair & Arxiv & \cite{25adaptiveTestGeneration} \\
6 & An Initial Investigation of ChatGPT Unit Test Generation   Capability & 2023 & Unit test case generation & ChatGPT & Zero-shot learning & Code & Pure LLM & SAST 2023 & \cite{150AnInitialInvestigation} \\
7 & Automated Test Case Generation Using Code Models and Domain   Adaptation & 2023 & Unit test case generation & CodeT5; LLaMA-2 & Pre-training and/or Fine-tuning & Code & Syntactic repair & Arxiv & \cite{119AutomatedTestCase} \\
8 & Automatic Generation of Test Cases based on Bug Reports: a   Feasibility Study with Large Language Models & 2023 & Unit test case generation & CodeGPT; ChatGPT & Pre-training and/or Fine-tuning & Bug description & Pure LLM & Arxiv & \cite{110AutomaticGenerationOf} \\
9 & Can Large Language Models Write Good Property-Based Tests? & 2023 & Unit test case generation & GPT-4 & Zero-shot learning & Code; Others & Pure LLM & Arxiv & \cite{121CanLargeLanguage} \\
10 & CAT-LM Training Language Models on Aligned Code And Tests & 2023 & Unit test case generation & GPT-neox & Pre-training and/or Fine-tuning & Code & Pure LLM & ASE 2023 & \cite{111CatLmTraining} \\
11 & ChatGPT vs SBST: A Comparative Assessment of Unit Test Suite   Generation & 2023 & Unit test case generation & ChatGPT & Zero-shot learning & Code & Pure LLM & Arxiv & \cite{123ChatgptVsSbst} \\
12 & ChatUniTest: a ChatGPT-based Automated Unit Test Generation   Tool & 2023 & Unit test case generation & ChatGPT & Zero-shot learning & Code & Syntactic repair & Arxiv & \cite{63chatUniTest} \\
13 & CODAMOSA: Escaping Coverage Plateaus in Test Generation with   Pre-trained Large Language Models & 2023 & Unit test case generation & Codex & Zero-shot learning & Code & Mutation testing; Program analysis & ICSE 2023 & \cite{55codamosaTestGeneration} \\
14 & Effective Test Generation Using Pre-trained Large Language   Models and Mutation Testing & 2023 & Unit test case generation & Codex & Few-shot learning; Zero-shot learning & Code & Mutation testing; Syntactic repair & Arxiv & \cite{117EffectiveTestGeneration} \\
15 & Exploring the Effectiveness of Large Language Models in   Generating Unit Tests & 2023 & Unit test case generation & CodeGen; Codex; ChatGPT & Zero-shot learning & Code & Syntactic repair & Arxiv & \cite{66generatingUnitTests} \\
16 & How Well does LLM Generate Security Tests? & 2023 & Unit test case generation & ChatGPT & Few-shot learning & Code & Pure LLM & Arxiv & \cite{112HowWellDoes} \\
17 & No More Manual Tests? Evaluating and Improving ChatGPT for   Unit Test Generation & 2023 & Unit test case generation & ChatGPT & Zero-shot learning & Code; Error information & Program analysis & Arxiv & \cite{64noUnitTest} \\
18 & Prompting Code Interpreter to Write Better Unit Tests on   Quixbugs Functions & 2023 & Unit test case generation & GPT-4 & Few-shot learning & Code & Pure LLM & Arxiv & \cite{113PromptingCodeInterpreter} \\
19 & Reinforcement Learning from Automatic Feedback for   High-Quality Unit Test Generation & 2023 & Unit test case generation & Codex & Pre-training and/or Fine-tuning & Code & Program analysis, Reinforcement learning & Arxiv & \cite{steenhoek2023reinforcement} \\
20 & Unit Test Generation using Generative AI: A Comparative   Performance Analysis of Autogeneration Tools & 2023 & Unit test case generation & ChatGPT & Zero-shot learning & Code & Pure LLM & Arxiv & \cite{bhatia2023unit} \\
21 & Generating Accurate Assert Statements for Unit Test Cases   Using Pretrained Transformers & 2023 & Test oracle generation & BART & Pre-training and/or Fine-tuning & Code & Pure LLM & AST 2022 & \cite{39assertStatements} \\
22 & Learning Deep Semantics for Test Completion & 2023 & Test oracle generation & CodeT5 & Pre-training and/or Fine-tuning & Code & Statistic analysis & ICSE 2023 & \cite{24semanticsTestCompletion} \\
23 & Using Transfer Learning for Code-Related Tasks & 2022 & Test oracle generation; Program repair & T5 & Pre-training and/or Fine-tuning & Code & Pure LLM & TSE 2022 & \cite{80UsingTransferLearning} \\
24 & Retrieval-Based Prompt Selection for Code-Related Few-Shot   Learning & 2023 & Test oracle generation; Program repair & Codex & Few-shot learning & Code & Pure LLM & ICSE 2023 & \cite{56retrievalPromptSelection} \\

\end{tabular}
\end{sidewaystable*}
\begin{sidewaystable*}[!t]
\footnotesize
\begin{tabular}{p{0.3cm}|p{5.5cm}|p{0.5cm}|p{2.5cm}|p{2.5cm}|p{3cm}|p{2cm}|p{2.5cm}|p{1.5cm}|p{0.5cm}}
\hline
\textbf{ID} & \textbf{Paper title} & \textbf{Year} & \textbf{Topic} & \textbf{Involved  LLM} & \textbf{How LLM is used} & \textbf{Input to LLM} & \textbf{How LLM integrated} & \textbf{Venue} & \textbf{Ref} \\ \hline
25 & Automated Conformance Testing for JavaScript Engines via Deep   Compiler Fuzzing & 2021 & System test input generation & GPT-2 & Pre-training and/or Fine-tuning & Code & Differential testing; Program analysis & PLDI 2021 & \cite{14conformanceTesting} \\
26 & Fill in the Blank: Context-aware Automated Text Input   Generation for Mobile GUI Testing & 2022 & System test input generation & GPT-3 & Pre-training and/or Fine-tuning & View hierarchy file of UI & Pure LLM & ICSE 2023 & \cite{26fillBlank} \\
27 & Large Language Models are Pretty Good Zero-Shot Video Game Bug   Detectors & 2022 & System test input generation & InstructGPT & Chain-of-Thought; Zero-shot learning & Others & Pure LLM & Arxiv & \cite{7gameBugDetectors} \\
28 & Slgpt: Using Transfer Learning to Directly Generate Simulink   Model Files and Find Bugs in the Simulink Toolchain & 2022 & System test input generation & GPT-2 & Pre-training and/or Fine-tuning & Others & Formal method & EASE 2021 & \cite{12slgptSimulink} \\
29 & Augmenting Greybox Fuzzing with Generative AI & 2023 & System test input generation & ChatGPT & Few-shot learning & Code & Pure LLM & Arxiv & \cite{141AugmentingGreyboxFuzzing} \\
30 & Automated Test Case Generation Using T5 and GPT-3 & 2023 & System test input generation & GPT-3; T5 & Pre-training and/or Fine-tuning; Zero-shot learning & NL specification & Pure LLM & ICACCS 2023 & \cite{143AutomatedTestCase} \\
31 & Automating GUI-based Software Testing with GPT-3 & 2023 & System test input generation & GPT-3 & Pre-training and/or Fine-tuning & View hierarchy file of UI & Pure LLM & ICSTW 2023 & \cite{84automatingGUI} \\
32 & AXNav: Replaying Accessibility Tests from Natural Language & 2023 & System test input generation & GPT-4 & Chain-of-Thought & View hierarchy file of UI & Pure LLM & Arxiv & \cite{159AxnavReplayingAccessibility} \\
33 & Can ChatGPT Advance Software Testing Intelligence? An   Experience Report on Metamorphic Testing & 2023 & System test input generation & ChatGPT & Zero-shot learning & Others & Pure LLM & Arxiv & \cite{106CanChatgptAdvance} \\
34 & Efficient Mutation Testing via Pre-Trained Language Models & 2023 & System test input generation & CodeBert & Zero-shot learning & Code & Mutation testing & Arxiv & \cite{30mutationTesting} \\
35 & Large Language Models are Edge-Case Generators:Crafting   Unusual Programs for Fuzzing Deep Learning Libraries & 2023 & System test input generation & Codex & Chain-of-Thought; Pre-training and/or Fine-tuning; Zero-shot   learning; Few-shot  learning & Code & Differential testing & ICSE 2024 & \cite{19fuzzDeepLearningLibraries} \\
36 & Large Language Models are Zero Shot Fuzzers: Fuzzing Deep   Learning Libraries via Large Language Models & 2023 & System test input generation & Codex; InCoder & Zero-shot learning & Code & Mutation testing; Differential testing & ISSTA 2023 & \cite{13fuzzDeepLearningLibraries} \\
37 & Large Language Models for Fuzzing Parsers (Registered Report) & 2023 & System test input generation & GPT-4 & Few-shot learning & NL specification & Pure LLM & FUZZING 2023 & \cite{152LargeLanguageModels} \\
38 & LLM for Test Script Generation and Migration: Challenges,   Capabilities, and Opportunities & 2023 & System test input generation & ChatGPT & Zero-shot learning & View hierarchy file of UI & Pure LLM & Arxiv & \cite{115LlmForTest} \\
39 & Make LLM a Testing Expert: Bringing Human-like Interaction to   Mobile GUI Testing via Functionality-aware Decisions & 2023 & System test input generation & GPT-3 & Zero-shot learning & View hierarchy file of UI & Natural language processing & ICSE 2024 & \cite{60GUITesting} \\
40 & PentestGPT: An LLM-empowered Automatic Penetration Testing   Tool & 2023 & System test input generation & ChatGPT; GPT-4; LaMDA & Chain-of-Thought; Few-shot learning & NL specification & Pure LLM & Arxiv & \cite{120PentestgptAnLlm} \\
41 & SMT Solver Validation Empowered by Large Pre-Trained Language   Models & 2023 & System test input generation & GPT-2 & Pre-training and/or Fine-tuning & Code & Differential testing & ASE 2023 & \cite{204SmtSolverValidation} \\
42 & TARGET: Automated Scenario Generation from Traffic Rules for   Testing Autonomous Vehicles & 2023 & System test input generation & GPT-3 & Zero-shot learning & Others & Scenario testing & Arxiv & \cite{62targetTestGeneration} \\
43 & Testing the Limits: Unusual Text Inputs Generation for Mobile   App Crash Detection with Large Language Model & 2023 & System test input generation & ChatGPT & Few-shot learning & View hierarchy file of UI & Pure LLM & ICSE 2024 & \cite{107TestingTheLimits} \\
44 & Understanding Large Language Model Based Fuzz Driver   Generation & 2023 & System test input generation & ChatGPT; GPT-4 & Few-shot learning; Zero-shot learning & Code; Others & Pure LLM & Arxiv & \cite{140UnderstandingLargeLanguage} \\
45 & Universal Fuzzing via Large Language Models & 2023 & System test input generation & GPT-4; StarCoder & Few-shot learning; Automatic prompt & Code & Mutation testing & ICSE 2024 & \cite{166Fuzz4allUniversalFuzzing} \\
46 & Variable Discovery with Large Language Models for Metamorphic   Testing of Scientific Software & 2023 & System test input generation & GPT-j & Zero-shot learning & Others & Pure LLM & SANER 2023 & \cite{82variablediscovery} \\

\end{tabular}
\end{sidewaystable*}
\begin{sidewaystable*}[!t]
\footnotesize
\begin{tabular}{p{0.3cm}|p{5.5cm}|p{0.5cm}|p{2.5cm}|p{2.5cm}|p{3cm}|p{2cm}|p{2.5cm}|p{1.5cm}|p{0.5cm}}
\hline
\textbf{ID} & \textbf{Paper title} & \textbf{Year} & \textbf{Topic} & \textbf{Involved  LLM} & \textbf{How LLM is used} & \textbf{Input to LLM} & \textbf{How LLM integrated} & \textbf{Venue} & \textbf{Ref} \\ \hline
47 & White-box Compiler Fuzzing Empowered by Large Language Models & 2023 & System test input generation & GPT-4; StarCoder & Few-shot learning & Code & Pure LLM & Arxiv & \cite{175WhiteBoxCompiler} \\
48 & Itiger: an Automatic Issue Title Generation Tool & 2022 & Bug analysis & BART & Pre-training and/or Fine-tuning & Bug description & Pure LLM & FSE 2022 & \cite{1itigerIssueTitle} \\
49 & CrashTranslator: Automatically Reproducing Mobile Application   Crashes Directly from Stack Trace & 2023 & Bug analysis & ChatGPT & Pre-training and/or Fine-tuning & Bug description & Reinforcement learning & ICSE 2024 & \cite{173CrashtranslatorAutomaticallyReproducing} \\
50 & Cupid: Leveraging ChatGPT for More Accurate Duplicate Bug   Report Detection & 2023 & Bug analysis & ChatGPT & Zero-shot learning & Bug description & Statistic analysis & Arxiv & \cite{132CupidLeveragingChatgpt} \\
51 & Employing Deep Learning and Structured Information Retrieval   to Answer Clarification Questions on Bug Reports & 2023 & Bug analysis & CodeT5 & Zero-shot learning & Bug description & Statistic analysis & Arxiv & \cite{134EmployingDeepLearning} \\
52 & Explaining Software Bugs Leveraging Code Structures in Neural   Machine Translation & 2023 & Bug analysis & CodeT5 & Pre-training and/or Fine-tuning & Code & Program analysis & ICSE 2023 & \cite{5explainingSoftwareBugs} \\
53 & Prompting Is All Your Need: Automated Android Bug Replay with   Large Language Models & 2023 & Bug analysis & ChatGPT & Few-shot learning; Chain-of-Thought & Bug description & Pure LLM & ICSE 2024 & \cite{161PromptingIsAll} \\
54 & Still Confusing for Bug-Component Triaging? Deep Feature   Learning and Ensemble Setting to Rescue & 2023 & Bug analysis & CodeT5 & Pre-training and/or Fine-tuning & Bug description & Statistic analysis & ICPC 2023 & \cite{145StillConfusingFor} \\
55 & Detect-Localize-Repair: A Unified Framework for Learning to   Debug with CodeT5 & 2022 & Debug & CodeT5 & Pre-training and/or Fine-tuning & Code & Pure LLM & EMNLP 2022 & \cite{31detectLocalizeRepair} \\
56 & Large Language Models are Few-shot Testers: Exploring   LLM-based General Bug Reproduction & 2022 & Debug & Codex & Few-shot learning & Bug description & Program analysis; Statistic analysis & ICSE 2023 & \cite{9generalBugReproduction} \\
57 & A Preliminary Evaluation of LLM-Based Fault Localization & 2023 & Debug & ChatGPT & Few-shot learning & Code & Pure LLM & Arxiv & \cite{163APreliminaryEvaluation} \\
58 & Addressing Compiler Errors: Stack Overflow or Large Language   Models? & 2023 & Debug & ChatGPT; GPT-4 & Zero-shot learning & Error information & Pure LLM & Arxiv & \cite{171AddressingCompilerErrors} \\
59 & Can LLMs Demystify Bug Reports? & 2023 & Debug & ChatGPT & Zero-shot learning & Bug description & Pure LLM & Arxiv & \cite{130CanLlmsDemystify} \\
60 & Dcc --help: Generating Context-Aware Compiler Error   Explanations with Large Language Models & 2023 & Debug & ChatGPT & Zero-shot learning & Code; Error information & Pure LLM & SIGCSE 2024 & \cite{164DccHelpGenerating} \\
61 & Explainable Automated Debugging via Large Language   Model-driven Scientific Debugging & 2023 & Debug & CodeGen; Codex; ChatGPT & Self-consistency; Zero-shot learning & Code & Pure LLM & Arxiv & \cite{54explainableDebugging} \\
62 & Large Language Models for Test-Free Fault Localization & 2023 & Debug & CodeGen & Pre-training and/or Fine-tuning & Code & Pure LLM & ICSE 2024 & \cite{168LargeLanguageModels} \\
63 & Large Language Models in Fault Localisation & 2023 & Debug & ChatGPT; GPT-4 & Zero-shot learning & Code; Error information & Pure LLM & Arxiv & \cite{136LargeLanguageModels} \\
64 & LLM4CBI: Taming LLMs to Generate Effective Test Programs for   Compiler Bug Isolation & 2023 & Debug & ChatGPT & Zero-shot learning & Code & Mutation testing; Reinforcement learning & Arxiv & \cite{122Llm4cbiTamingLlms} \\
65 & Nuances are the Key: Unlocking ChatGPT to Find   Failure-Inducing Tests with Differential Prompting & 2023 & Debug & ChatGPT & Zero-shot learning & Code & Differential testing & ASE 2023 & \cite{67failureTestCases} \\
66 & Teaching Large Language Models to Self-Debug & 2023 & Debug & Codex; ChatGPT; GPT-4; StarCoder & Few-shot learning & Code & Pure LLM & Arxiv & \cite{139TeachingLargeLanguage} \\
67 & A study on Prompt Design, Advantages and Limitations of   ChatGPT for Deep Learning Program Repair & 2023 & Debug; Program repair & ChatGPT & Zero-shot learning & Code & Pure LLM & Arxiv & \cite{52promptDesign} \\
68 & Examining Zero-Shot Vulnerability Repair with Large Language   Models & 2021 & Program repair & Codex & Zero-shot learning & Code; Bug description & Pure LLM & SP 2023 & \cite{35vulnerabilityRepair} \\
69 & Automated Repair of Programs from Large Language Models & 2022 & Program repair & Codex & Zero-shot learning & Code & Pure LLM & ICSE 2023 & \cite{34automatedRepair} \\
70 & Fix Bugs with Transformer through a Neural-Symbolic Edit   Grammar & 2022 & Program repair & CodeGPT & Pre-training and/or Fine-tuning & Code & Pure LLM & Arxiv & \cite{10neuralSymbolic} \\
71 & Practical Program Repair in the Era of Large Pre-trained   Language Models & 2022 & Program repair & GPT-3; Codex; CodeT5; InCoder & Few-shot learning; Zero-shot learning & Code & Statistic analysis & ICSE 2023 & \cite{33practicalProgramRepair} \\

\end{tabular}
\end{sidewaystable*}
\begin{sidewaystable*}[!t]
\footnotesize
\begin{tabular}{p{0.3cm}|p{5.5cm}|p{0.5cm}|p{2.5cm}|p{2.5cm}|p{3cm}|p{2cm}|p{2.5cm}|p{1.5cm}|p{0.5cm}}
\hline
\textbf{ID} & \textbf{Paper title} & \textbf{Year} & \textbf{Topic} & \textbf{Involved  LLM} & \textbf{How LLM is used} & \textbf{Input to LLM} & \textbf{How LLM integrated} & \textbf{Venue} & \textbf{Ref} \\ \hline
72 & Repairing Bugs in Python Assignments Using Large Language   Models & 2022 & Program repair & Codex & Few-shot learning; Zero-shot learning & Code; Error information & Program analysis & Arxiv & \cite{8bugsPythonAssignments} \\
73 & Towards JavaScript Program Repair with Generative Pre-trained   Transformer (GPT-2) & 2022 & Program repair & GPT-2 & Pre-training and/or Fine-tuning & Code & Pure LLM & APR 2022 & \cite{38javaScriptProgram} \\
74 & An Analysis of the Automatic Bug Fixing Performance of ChatGPT & 2023 & Program repair & ChatGPT & Zero-shot learning & Code; Error information & Pure LLM & APR 2023 & \cite{4analysisBugFixing} \\
75 & An Empirical Study on Fine-Tuning Large Language Models of   Code for Automated Program Repair & 2023 & Program repair & PLBART; CodeT5; UniXCoder & Pre-training and/or Fine-tuning & Code & Pure LLM & ASE 2023 & \cite{206AnEmpiricalStudy} \\
76 & An Evaluation of the Effectiveness of OpenAI's ChatGPT for   Automated Python Program Bug Fixing using QuixBugs & 2023 & Program repair & ChatGPT & Zero-shot learning & Code & Pure LLM & iSemantic 2023 & \cite{144AnEvaluationOf} \\
77 & An Extensive Study on Model Architecture and Program   Representation in the Domain of Learning-based Automated Program Repair & 2023 & Program repair & T5; CodeT5 & Pre-training and/or Fine-tuning & Code & Pure LLM & APR 2023 & \cite{147AnExtensiveStudy} \\
78 & Can OpenAI's Codex Fix Bugs? An Evaluation on QuixBugs & 2023 & Program repair & Codex & Few-shot learning; Zero-shot learning & Code & Pure LLM & APR 2022 & \cite{44codexQuixBugs} \\
79 & CIRCLE: Continual Repair Across Programming Languages & 2023 & Program repair & T5 & Pre-training and/or Fine-tuning & Code & Pure LLM & ISSTA 2022 & \cite{40circleContinualRepair} \\
80 & Coffee: Boost Your Code LLMs by Fixing Bugs with Feedback & 2023 & Program repair & CodeLLAMA & Pre-training and/or Fine-tuning & Code & Pure LLM & Arxiv & \cite{moon2023coffee} \\
81 & Copiloting the Copilots: Fusing Large Language Models with   Completion Engines for Automated Program Repair & 2023 & Program repair & CodeT5; InCoder & Zero-shot learning & Code & Statistic analysis & FSE 2023 & \cite{201CopilotingTheCopilots} \\
82 & Domain Knowledge Matters: Improving Prompts with Fix Templates   for Repairing Python Type Errors & 2023 & Program repair & CodeT5 & Pre-training and/or Fine-tuning & Code & Program analysis & ICSE 2024 & \cite{DomainKnowledge} \\
83 & Enhancing Genetic Improvement Mutations Using Large Language   Models & 2023 & Program repair & GPT-4 & Zero-shot learning & Code & Pure LLM & SSBSE 2023 & \cite{167EnhancingGeneticImprovement} \\
84 & FixEval: Execution-based Evaluation of Program Fixes for   Programming Problems & 2023 & Program repair & CodeT5; PLBART & Pre-training and/or Fine-tuning & Code & Pure LLM & APR 2023 & \cite{149FixevalExecutionBased} \\
85 & Fixing Hardware Security Bugs with Large Language Models & 2023 & Program repair & Codex; CodeGen & Few-shot learning; Zero-shot learning & Code; Bug description & Pure LLM & Arxiv & \cite{36securityBugs} \\
86 & Fixing Rust Compilation Errors using LLMs & 2023 & Program repair & ChatGPT; GPT-4 & Zero-shot learning & Code & Pure LLM & Arxiv & \cite{104FixingRustCompilation} \\
87 & Framing Program Repair as Code Completion & 2023 & Program repair & CodeGPT & Zero-shot learning & Code & Pure LLM & ICSE 2022 & \cite{41repairAsCompletion} \\
88 & Frustrated with Code Quality Issues? LLMs can Help! & 2023 & Program repair & ChatGPT; GPT-4 & Zero-shot learning & Code & Pure LLM & Arxiv & \cite{135FrustratedWithCode} \\
89 & GPT-3-Powered Type Error Debugging: Investigating the Use of   Large Language Models for Code Repair & 2023 & Program repair & GPT-3 & Zero-shot learning & Code & Program analysis & SLE 2023 & \cite{151Gpt3Powered} \\
90 & How Effective Are Neural Networks for Fixing Security   Vulnerabilities & 2023 & Program repair & Codex; CodeGen; CodeT5; PLBART; InCoder & Pre-training and/or Fine-tuning; Zero-shot learning & Code & Pure LLM & ISSTA 2023 & \cite{57howEffective} \\
91 & Impact of Code Language Models on Automated Program Repair & 2023 & Program repair & PLBART; CodeT5; CodeGen; InCoder & Pre-training and/or Fine-tuning; Zero-shot learning & Code & Pure LLM & ICSE 2023 & \cite{32impactLanguageModels} \\
92 & Inferfix: End-to-end Program Repair with LLMs & 2023 & Program repair & Codex & Few-shot learning; Pre-training and/or Fine-tuning & Code & Pure LLM & FSE 2023 & \cite{77inferfixProgramRepair} \\
93 & Keep the Conversation Going: Fixing 162 out of 337 bugs for   \$0.42 each using ChatGPT & 2023 & Program repair & ChatGPT & Few-shot learning & Code; Error information & Pure LLM & Arxiv & \cite{2keepConversationGoing} \\
94 & Neural Program Repair with Program Dependence Analysis and   Effective Filter Mechanism & 2023 & Program repair & CodeT5 & Pre-training and/or Fine-tuning & Code & Statistic analysis & Arxiv & \cite{71repairProgramDependence} \\

\end{tabular}
\end{sidewaystable*}
\begin{sidewaystable*}[!t]
\footnotesize
\begin{tabular}{p{0.3cm}|p{5.5cm}|p{0.5cm}|p{2.5cm}|p{2.5cm}|p{3cm}|p{2cm}|p{2.5cm}|p{1.5cm}|p{0.5cm}}
\hline
\textbf{ID} & \textbf{Paper title} & \textbf{Year} & \textbf{Topic} & \textbf{Involved  LLM} & \textbf{How LLM is used} & \textbf{Input to LLM} & \textbf{How LLM integrated} & \textbf{Venue} & \textbf{Ref} \\ \hline
95 & Out of Context: How important is Local Context in Neural   Program Repair? & 2023 & Program repair & CodeT5 & Pre-training and/or Fine-tuning & Code & Pure LLM & ICSE 2024 & \cite{prenner2023context} \\
96 & Pre-trained Model-based Automated Software Vulnerability   Repair: How Far are We? & 2023 & Program repair & CodeT5; UniXCoder; CodeGPT & Pre-training and/or Fine-tuning & Code & Pure LLM & IEEE TDSC & \cite{148PreTrainedModel} \\
97 & RAPGen: An Approach for Fixing Code Inefficiencies in   Zero-Shot & 2023 & Program repair & Codex & Few-shot learning; Chain-of-Thought & Code & Pure LLM & Arxiv & \cite{154RapgenAnApproach} \\
98 & RAP-Gen: Retrieval-Augmented Patch Generation with CodeT5 for   Automatic Program Repair & 2023 & Program repair & CodeT5 & Pre-training and/or Fine-tuning & Code & Statistic analysis & FSE 2023 & \cite{202RapGenRetrieval} \\
99 & STEAM: Simulating the InTeractive BEhavior of ProgrAMmers for   Automatic Bug Fixing & 2023 & Program repair & ChatGPT & Zero-shot learning & Code & Pure LLM & Arxiv & \cite{102SteamSimulatingThe} \\
100 & Towards Generating Functionally Correct Code Edits from   Natural Language Issue Descriptions & 2023 & Program repair & Codex; ChatGPT & Few-shot learning; Zero-shot learning; Chain-of-Thought & Code; Bug description & Pure LLM & Arxiv & \cite{68codeEdits} \\
101 & VulRepair: a T5-based Automated Software Vulnerability Repair & 2023 & Program repair & T5 & Pre-training and/or Fine-tuning & Code & Pure LLM & FSE 2022 & \cite{43vulRepairVulnerabilityRepair} \\
102 & What Makes Good In-Context Demonstrations for Code   Intelligence Tasks with LLMs? & 2023 & Program repair & Codex; ChatGPT & Few-shot learning & Code & Pure LLM & ASE 2023 & \cite{205WhatMakesGood} \\

\end{tabular}
\end{sidewaystable*}
\end{document}